\documentclass[journal]{IEEEtran}

\usepackage{amsmath,amssymb,amsfonts}
\usepackage{algorithmic}
\usepackage{graphicx}
\usepackage{textcomp}
\usepackage{xcolor}
\usepackage{pdfpages}
\usepackage{subcaption}
\usepackage[T1]{fontenc}
\usepackage{multirow}

\usepackage{booktabs}
\usepackage[displaymath, mathlines,switch]{lineno}
\begin{document}
\title{Time-Localized Parametric Decomposition of Respiratory Airflow for Sub-Breath Analysis}
\author{Victoria Ribeiro Rodrigues, Paul W. Davenport, and Nicholas J. Napoli
\thanks{This work was supported by research funding from the Office of Naval Research (grant number N00014-22-1-2653)}
\thanks{Victoria Ribeiro Rodrigues and Nicholas J. Napoli are with the Human Informatics and Predictive Performance Optimization (HIPPO) Laboratory, Department of Electrical and Computer Engineering, University of Florida, Gainesville, FL 32611 USA (e-mail: \{victoria.ribeiro, n.napoli\}@ufl.edu).}
\thanks{Paul W. Davenport is with the Department of Physiological Sciences, University of Florida, Gainesville, FL 32611 USA (e-mail: pdavenpo@ufl.edu).}
}

\date{\today}
\markboth{Journal of \LaTeX\ Class Files,~Vol.~14, No.~8, August~2015}%
{Shell \MakeLowercase{\textit{et al.}}: Bare Demo of IEEEtran.cls for IEEE Journals}



\maketitle

\begin{abstract}
Respiratory airflow signals provide critical insight into breathing mechanics, yet conventional analysis methods remain limited in their ability to characterize the internal structure of individual breaths. Traditional approaches treat airflow as a quasi-periodic signal and rely on global descriptors such as tidal volume or peak flow, obscuring sub-breath events that reflect neuromuscular coordination and compensatory breathing strategies. This study introduces a parametric framework for decomposing inspiratory airflow into a small number of time-localized components with explicit amplitude, onset time, and duration parameters. Unlike spectral or data-adaptive methods, the proposed approach employs physiologically grounded basis functions, Half-Sine, Gaussian, and Beta, to represent intrabreath waveform morphology through constrained nonlinear optimization. Evaluation across 8,276 breaths demonstrates high reconstruction accuracy (mean squared error $<$ 0.001 for four-component models) and robust parameter precision under moderate noise. Component-derived features describing sub-breath timing and coordination improved classification of cognitive fatigue states arising from cognitive-respiratory competition by up to 30.7\% in Matthews correlation coefficient compared with classical respiratory metrics. These results establish that modeling airflow as a sum of parameterized, time-localized primitives provides an interpretable and precise foundation for quantifying intrabreath organization, compensatory breathing dynamics, and respiratory motor control adaptation under cognitive-respiratory dual-task demands.

\end{abstract}

\section{Introduction}
Airflow signals are widely used in clinical and research settings because they are easy to collect and provide a non-invasive view of breathing mechanics~\cite{vicarioNoninvasive2016, grayRespiration1951}. The goal of this work is to capture and parameterize the internal subphases of inspiratory airflow, enabling analysis of how timing and coordination within a single breath adapt under changing physiological or cognitive demands. Despite decades of research in respiratory signal analysis, existing methods remain fundamentally limited in their ability to describe this internal structure. Airflow is commonly treated as a low-frequency, quasi-periodic signal, and individual inspirations are summarized using global descriptors such as tidal volume, inspiratory duration, or peak flow~\cite{ciolekAutomated2015}. Although respiration exhibits an overall rhythmic structure, the airflow waveform within each respiratory cycle is highly nonstationary and shaped by dynamic interactions among neural drive, muscular recruitment, and respiratory mechanics~\cite{fiammaEffects2007, hessNeural2013, onimaruNovel2003, napoliCharacterizing2018}. As a result, potentially meaningful sub-breath events, such as changes in rise dynamics, inflections, or phase-specific contributions, are obscured when each breath is reduced to a single set of scalar metrics.

Understanding these sub-breath events is essential because breathing adapts continuously to physiological and environmental demands through multiple compensatory mechanisms. In response to external load, disease, hypoxia, or fatigue, neural drive modifies muscle recruitment and breathing patterns to maintain effective ventilation and gas exchange while balancing mechanical and metabolic constraints~\cite{otisEffect1949, zechmanEffect1977, harberPhysiologic1982, grassmannRespiratory2016a, romerExerciseinduced2008, mckenzieRespiratory2012, milic-emiliRelationship2011, RODRIGUES2022}. These compensatory adjustments are often expressed as changes in airflow waveform morphology, such as altered rise slopes, inflections, or asymmetries, reflecting shifts in the timing and coordination of inspiratory effort even when global timing and magnitude measures remain relatively stable~\cite{bachyprogram1986, abboudFrequency1986, benchetritIndividuality1989, satoMethods2001}.

Each breath is not a single continuous event but rather the result of overlapping activations of multiple respiratory muscle groups, each contributing a distinct portion of the inspiratory airflow. These sub-events reflect time-varying neural recruitment and redistribution of mechanical load throughout inspiration~\cite{fiammaEffects2007, hessNeural2013, napoliCharacterizing2018}. Representing these components explicitly through parameters such as onset time, duration, and amplitude enables quantification of how inspiratory contributions evolve under different physiological or cognitive demands. A parametric decomposition of the airflow waveform therefore provides a structured description of breathing control, linking observed waveform morphology to the internal organization of respiratory effort.

\subsection{Prior Work}

Analyses of respiratory airflow have traditionally focused on characterizing global breathing patterns over time rather than the internal structure of individual breaths. Most approaches treat airflow as a smooth, quasi-periodic signal and use frequency- or time–frequency–domain methods to describe its rhythmic and energetic properties. This perspective has provided important insight into ventilatory control and clinical classification but offers limited access to the short, transient events that shape the morphology of each inspiratory cycle.

\textbf{Spectral Analysis Approaches.}
Fourier-based techniques have long been used to quantify the frequency content of respiratory flow. Early work applied power spectral density (PSD) and Fourier descriptors to identify respiratory disorders~\cite{abboudFrequency1986}. More recent studies have used spectral and PSD-derived features from airflow signals to classify sleep apnea severity in adults and children~\cite{jimenez-garciaAssessment2020, gardeBreathing2010}. Recent advances have further extended respiratory signal processing using spectral and time-frequency methods for sleep monitoring~\cite{8688505}, disease detection~\cite{10904160}, and respiratory event classification~\cite{9373894}. These analyses describe dominant oscillatory components and overall rhythmic stability but provide aggregate measures across an analysis window. As a result, they primarily capture global frequency behavior rather than time-localized changes within a single breath.

The Fourier Transform models the signal as a sum of infinite-duration sinusoids~\cite{bracewell2000, cohen1995}. When applied to short inspiratory segments, its frequency resolution is inherently limited by signal duration, producing energy distributions that represent average content over the entire cycle. Windowed variants such as the Short-Time Fourier Transform (STFT)~\cite{allenShort1977, gaborTheory1946, ciolekanalysis2010} introduce temporal localization at the cost of frequency resolution~\cite{cohen1995}. These approaches are therefore well-suited to assessing the rhythm and overall variability of a long sequence of breaths, but not designed to resolve distinct, time-localized airflow features.

\textbf{Wavelet-Based Approaches.}
Wavelet transforms offer improved temporal adaptability and are commonly used in sleep apnea research involving airflow signals. Both discrete and continuous wavelet transforms have been used to extract multiscale features for respiratory event detection~\cite{emintaglukClassification2010, barroso-garciaWavelet2021}. Wavelet coefficients provide a joint time–scale representation that captures irregularities and asymmetries across multiple resolutions, supporting robust classification when combined with oximetry or pressure data~\cite{yadavMachine2020,kimSignal2001}. However, the resulting coefficients represent distributions of signal energy rather than explicit waveform components, which makes it difficult to interpret them in terms of physiological events or parameters such as amplitude, onset, or duration~\cite{mallattheory1989, daubechiesOrthonormal1988}. Wavelet-based approaches have been extensively applied for respiratory sound classification~\cite{9372748, 9310306}, demonstrating their utility for energy-based feature extraction, though the resulting representations remain challenging to interpret in terms of specific temporal events within individual breaths.

\textbf{Data-Driven and Dictionary-Based Approaches.}
Beyond analytic transforms, adaptive methods such as Matching Pursuit (MP) and K-SVD decompositions have been applied to other biosignals but only rarely to airflow~\cite{sommermeyerDetection}. These techniques reconstruct a signal by selecting or learning atoms from a dictionary, enabling flexible representation of transient patterns. In practice, however, the learned or chosen atoms describe statistical structure in the data and are not explicitly constrained by physiological or morphological models of breathing. Related approaches using empirical mode decomposition (EMD) have been applied to respiratory signals to isolate dominant oscillatory components~\cite{salisburyRapid2007, robertsonEMD2007}. EMD and ensemble EMD (EEMD) have been used to extract respiratory rate from photoplethysmographic signals~\cite{7873222, 7101812} and to differentiate normal from adventitious respiratory sounds~\cite{7021891}. While these decompositions provide adaptive flexibility, the resulting intrinsic modes vary in frequency and phase across realizations, making cross-subject or cross-condition interpretation challenging. Moreover, as demonstrated in~\cite{9310306}, hybrid approaches combining EMD with continuous wavelet transforms still produce scalogram-based energy representations rather than explicit temporal parameters.

\textbf{Morphological Analysis of Airflow Waveforms.}
Several studies have sought to characterize the overall morphology of the instantaneous airflow waveform using frequency-domain descriptors~\cite{bachyprogram1986, abboudFrequency1986, benchetritIndividuality1989}. These works established that waveform shape contains physiologically relevant information but relied on spectral analyses assuming stationarity within each breath. More recently, Symmetric Projection Attractor Reconstruction (SPAR) has been applied to respiratory airflow waveforms to describe their geometric morphology~\cite{serna-pascualNovel2023}. SPAR demonstrated that waveform shape alone can distinguish participant groups without relying on scalar timing metrics, underscoring the value of morphological analysis. Because SPAR constructs an attractor from multiple consecutive breaths, however, it emphasizes global geometry rather than resolving sub-events within individual inspirations or quantifying their contribution to total airflow. Other morphological approaches have analyzed respiratory cycle structure for event detection~\cite{9591404} and used attractor-based methods for COPD characterization~\cite{11283643}, demonstrating the clinical value of waveform shape analysis. However, these methods either focus on cycle segmentation and discrete event detection rather than continuous sub-breath decomposition, or construct attractors from multiple breaths to capture global dynamics rather than resolving intra-breath temporal organization.

\textbf{Interpretable Respiratory Variability Markers.}
Recent efforts have sought to develop interpretable markers that quantify respiratory pattern variability for clinical and cognitive state assessment. Studies have demonstrated that respiratory timing features provide sensitive indices of stress~\cite{7452349} and that interpretable respiratory variability markers can distinguish posttraumatic stress disorder symptoms~\cite{10521732}. These studies underscore the clinical value of features with clear temporal meaning, yet existing approaches rely on breath-to-breath variability or aggregate cycle statistics rather than structured decomposition of sub-breath coordination within individual inspirations.

\subsection{Challenges}

Despite decades of research in respiratory signal analysis, current methods are fundamentally limited in their ability to describe the internal structure of a single breath. Airflow signals are often treated as low-frequency, quasi-periodic time series, assuming that each cycle represents a stable repetition of the same underlying ventilatory pattern~\cite{abboudFrequency1986, benchetritBreathing2000}. This simplification is convenient for spectral or envelope-based analysis but physiologically inaccurate. Although respiration exhibits an overall rhythmic structure, the airflow waveform within each respiratory cycle is highly nonstationary and shaped by dynamic interactions among neural drive, mechanical load, and airway resistance~\cite{milic-emiliDrive1976}. The instantaneous flow rate can change abruptly as different muscle groups are recruited or as airway mechanics shift~\cite{fiammaEffects2007, hessNeural2013, napoliCharacterizing2018}. Consequently, airflow is not truly periodic but a transient signal whose morphology varies from one inspiration to the next~\cite{napoliCharacterizing2022}.

% challenge 1: Time–frequency methods cannot uniquely resolve transient, aperiodic subcomponents within a single inspiration.
From a signal processing standpoint, inspiratory airflow presents a fundamental mismatch with spectral and time–frequency analysis frameworks. A single inspiratory waveform is short in duration, typically no longer than two to three seconds, and does not consist of multiple oscillatory cycles but rather corresponds to at most a fraction of a single cycle~\cite{napoliCharacterizing2022}. Fourier-based representations, therefore, lack sufficient temporal support to define meaningful frequency content at the breath level~\cite{bracewell2000}. Time–frequency methods such as the STFT and Wavelet Transform attempt to localize spectral energy in time, but remain fundamentally constrained by the Heisenberg uncertainty principle, which imposes a lower bound on simultaneous time and frequency resolution~\cite{gaborTheory1946, cohen1995, mallattheory1989}. For a finite observation window of duration $T$, the minimum resolvable frequency separation is $\Delta f \approx 1/T$, and when the signal contains fewer than one oscillatory cycle, these limits prevent reliable estimation or localization of frequency components~\cite{cohen1995, gaborTheory1946}. As a result, short-lived, non-oscillatory sub-events within an inspiratory cycle, such as abrupt slope changes, inflections, or overlapping flow contributions, are spread across time–frequency representations and cannot be uniquely resolved or reconstructed as explicit waveform components with well-defined onset, duration, and amplitude.

\begin{figure*}[!ht]
\centering
\begin{subfigure}[b]{0.48\textwidth}
    \centering
    \includegraphics[width=\textwidth]{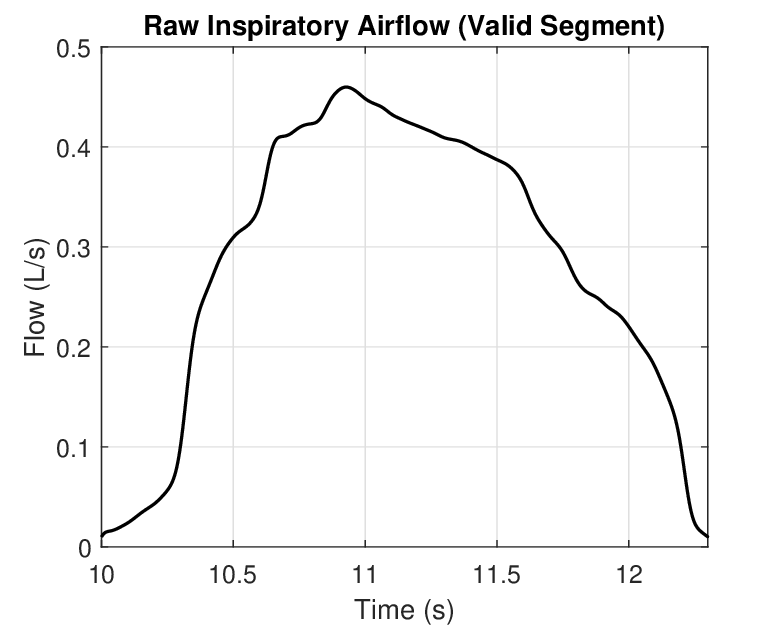}
    \caption{Raw airflow waveform (2.4~s inspiratory time) padded with 10~s of zeros on each side to reduce edge discontinuities.}
    \label{fig:airflow_raw}
\end{subfigure}
\hfill
\begin{subfigure}[b]{0.48\textwidth}
    \centering
    \includegraphics[width=\textwidth]{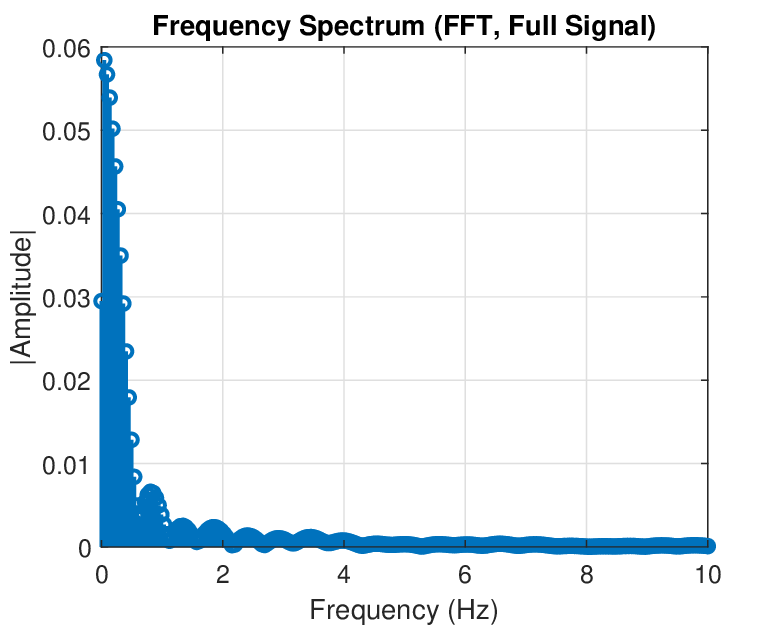}
    \caption{FFT amplitude spectrum of the zero-padded signal.}
    \label{fig:airflow_fft}
\end{subfigure}

\vspace{0.5em}

\begin{subfigure}[b]{0.48\textwidth}
    \centering
    \includegraphics[width=\textwidth]{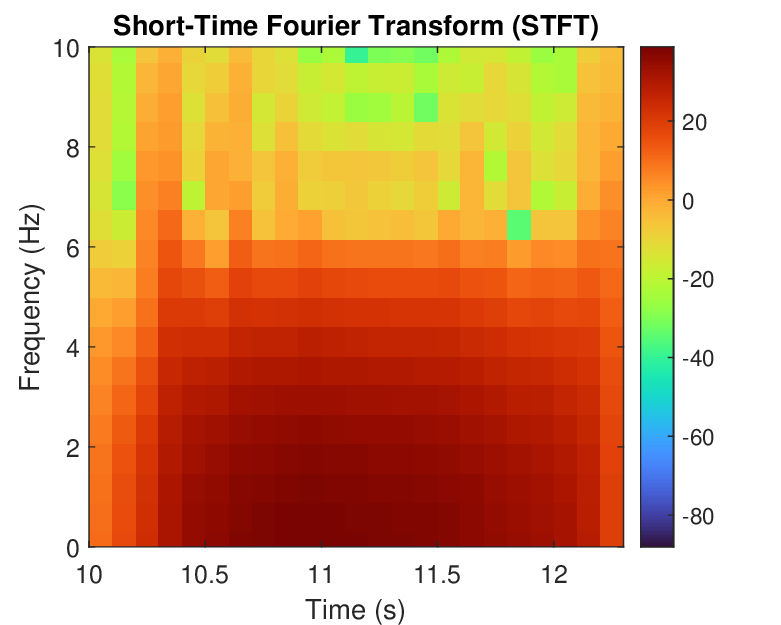}
    \caption{STFT spectrogram using a 0.3s Hamming window with 75\% overlap.}
    \label{fig:airflow_stft}
\end{subfigure}
\hfill
\begin{subfigure}[b]{0.48\textwidth}
    \centering
    \includegraphics[width=\textwidth]{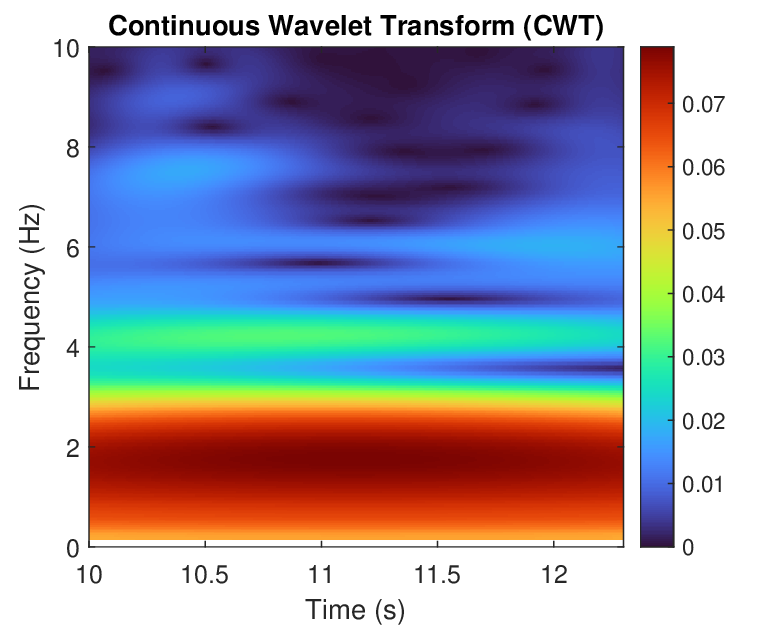}
    \caption{CWT Scalogram using Morse wavelet.}
    \label{fig:airflow_cwt}
\end{subfigure}

\caption[FFT, STFT, and CWT representations of a single inspiratory waveform.]{Comparison of spectral and time--frequency representations for a single inspiratory airflow waveform. (a) Raw inspiratory airflow segment (inspiratory time = 2.4 s), shown after zero-padding with 10 s of zeros on each side to reduce edge discontinuities in subsequent transforms. (b) Magnitude spectrum obtained via the FFT of the zero-padded signal. Zero-padding smooths spectral interpolation but does not increase true frequency resolution, which is fundamentally limited by the duration of the valid inspiratory segment. The minimum resolvable frequency separation is $\Delta f \approx 1/T$ (here $\approx$ 0.42 Hz), resulting in broad low-frequency energy concentration and poor discrimination of intrabreath structure. (c) STFT spectrogram computed using a 0.3 s Hamming window with 75\% overlap. Temporal localization is improved relative to the FFT, but at the expense of frequency resolution, as dictated by the time--frequency uncertainty principle. Sub-breath features are spread across time--frequency bins and cannot be isolated as distinct components. (d) Continuous wavelet transform (CWT) scalogram using a Morse wavelet. Across all representations, the short, non-oscillatory nature of a single inspiratory waveform prevents reliable frequency resolution and invertible decomposition at the breath level, motivating time-localized, component-based representations for intrabreath analysis.}
\label{fig:airflow_spectral_analysis}
\end{figure*}

Figure~\ref{fig:airflow_spectral_analysis} illustrates this limitation across several commonly used signal representations applied to a single inspiratory airflow waveform. Each method summarizes the signal in terms of coarse temporal or spectral energy distributions, but none preserves the fine temporal structure required to reconstruct the original inspiratory morphology. Because these representations project the waveform onto oscillatory or windowed bases with limited temporal support, transient sub-breath events are spread across time–frequency bins and are not uniquely identifiable. Once this projection is applied, the mapping is no longer invertible at the level of individual breaths, and the original waveform geometry cannot be recovered. If a signal cannot be decomposed with sufficient temporal and morphological resolution, accurate reconstruction is fundamentally impossible. This lack of reconstructability at the single-breath level constitutes a core limitation of existing approaches 

Even when adaptive methods extract apparent subcomponents, the derived parameters are not stable under small perturbations in the signal~\cite{huangempirical1998}. Because these algorithms lack explicit temporal regularization or joint parameter optimization, moderate noise can alter the number, order, and shape of recovered modes. The same underlying breathing pattern can therefore yield different decompositions across cycles or subjects. Without reproducible component definitions, parameters such as onset time, duration, and amplitude cannot be compared meaningfully or tracked across repeated measurements.

Data-adaptive methods such as Empirical Mode Decomposition (EMD), Matching Pursuit, and K-SVD were introduced to relax the periodicity assumptions of Fourier and wavelet analysis by adapting their basis functions directly to the signal. However, these approaches remain fundamentally limited because their adaptation is driven by statistical or reconstruction criteria rather than by an explicit temporal or morphological model of respiration. As a result, the recovered components are weakly constrained and lack a consistent correspondence to interpretable intrabreath structure. EMD produces intrinsic mode functions derived from local extrema whose number and shape depend on noise and boundary conditions. Matching Pursuit fits atoms sequentially without re-optimizing prior components, distorting temporal and amplitude relationships among overlapping events. K-SVD alternates between sparse coding and dictionary updates, yielding atoms that reflect aggregate signal statistics rather than stable, time-localized physiological processes. Across these methods, the central limitation is the absence of structural constraints that restrict the solution space to physiologically meaningful temporal configurations.

Finally, without a method that can identify intrabreath components with sufficient precision and with parameters that carry clear temporal meaning, such as amplitude, duration, and activation time, it is not possible to construct features that reflect how inspiratory airflow is organized and adapted within a single breath. Existing analyses, therefore, rely on global breath-level summaries or energy-based descriptors that collapse intrabreath structure, obscuring coordination, overlap, and phase-specific contributions. This lack of parameter-derived intrabreath features limits the ability to quantify compensatory breathing strategies and prevents systematic investigation of how airflow dynamics reorganize under changing physiological or cognitive demands.

Together, these challenges highlight fundamental gaps in the current literature that have limited explicit characterization of intrabreath airflow structure. First, existing spectral and time–frequency methods lack the resolution needed to represent and reconstruct sub-breath dynamics within a single inspiration. Second, available decomposition approaches do not provide sufficiently precise and repeatable component parameters to support meaningful comparison under noise or across repeated measurements. Third, even when intrabreath subcomponents are identified, current methods lack a structured and interpretable temporal parameterization from which features describing airflow organization and adaptation can be derived.

To address these gaps, this work is guided by three research questions:
\begin{itemize}
    \item \textbf{RQ1):} Can inspiratory airflow be reconstructed at the single-breath level using a small number of time-localized, interpretable components?
    \item \textbf{RQ2):} Does the proposed parametric decomposition yield precise and repeatable component parameters under moderate noise?
    \item \textbf{RQ3):} Do sub-breath features derived from the proposed representation capture meaningful adaptations in breathing dynamics that are not accessible through conventional breath-level measures?
\end{itemize}

To date, no single method provides a compact, interpretable description of the inspiratory airflow waveform in terms of a well-defined temporal basis with explicit amplitude, onset time, and duration, as required for meaningful within-breath analysis.

\subsection{Insights}

This work addresses the limitations identified above through a unified parametric framework that models each inspiratory breath as a sum of time-localized waveform components with interpretable parameters. The key insight is that intrabreath structure can be recovered not by increasing spectral resolution or adaptivity, but by explicitly modeling the temporal organization of airflow using a small number of constrained, physiologically plausible components. This approach balances flexibility, interpretability, and robustness, properties that are rarely achieved simultaneously in respiratory signal analysis.

First, to overcome the resolution and representational limitations of time–frequency methods, each inspiratory breath is represented using a finite set of simple, non-oscillatory basis functions that are explicitly localized in time. Because these components are defined over finite temporal support, they are well-suited to modeling short, non-periodic inspiratory waveforms. This formulation enables reconstruction of intrabreath morphology at the single-breath level, preserving transient features such as inflections, shoulders, and overlapping flow contributions that cannot be captured by spectral decomposition methods.

Second, to address the lack of parameter precision and reproducibility in existing decomposition approaches, the model enforces a compact and well-constrained parameterization. Each component is defined by amplitude, onset time, and duration, quantities with clear physical interpretations and bounded physiological ranges. All component parameters are estimated jointly through constrained nonlinear optimization, rather than sequentially or statistically. This global fitting strategy preserves internal timing relationships between overlapping components and yields repeatable parameter estimates under noise, even when the inverse problem is not strictly unique. Precision, therefore, becomes the foundation for interpretability and subsequent analysis.

Finally, building on this precise parameterization, the framework enables the construction of features that explicitly quantify how sub-breath events interact in time and amplitude within a single inspiration. Because each component has a well-defined onset, duration, and magnitude, interactions such as temporal overlap, relative timing, amplitude dominance, and phase-specific contributions can be measured directly. These features describe not only the presence of sub-breath events but also their coordination and redistribution across the inspiratory interval. Unlike data-adaptive methods that partition signal energy without consistent temporal semantics, the proposed representation provides a stable coordinate system for capturing intrabreath organization and compensatory breathing strategies that are not accessible through conventional breath-level metrics.

Together, these design choices establish a mathematically consistent and physiologically interpretable framework that resolves intrabreath structure, yields precise, reproducible parameters, and supports meaningful component-level and feature-based analysis of breathing dynamics across varying conditions.

\subsection{Contributions}

This study introduces a unified parametric framework for decomposing inspiratory airflow into a finite set of time-localized components with interpretable parameters. Unlike spectral or data-driven decompositions, the proposed approach employs physiologically grounded basis functions to represent intrabreath waveform morphology in terms of amplitude, onset time, duration, and shape. This formulation yields a compact and consistent representation of a single inspiration that is comparable across breaths, subjects, and experimental conditions.

The proposed framework enables accurate reconstruction of inspiratory airflow, produces precise and repeatable component parameters in the presence of noise, and exposes structured intrabreath subcomponents from which meaningful features can be derived. Using these component-level features, the method captures differences in breathing dynamics associated with fatigued and non-fatigued states that are not apparent from conventional breath-level measures. Together, these contributions demonstrate that modeling airflow as a sum of parameterized, time-localized primitives provides a practical and interpretable foundation for quantifying intrabreath organization and adaptive respiratory behavior.

\section{Methods}

\subsection{Model Overview}

Each inspiratory breath is modeled as a superposition of a finite number of time-localized waveform components.
Let \(x_n(t)\) denote the airflow signal of the \(n\)-th inspiratory breath, where \(t \in [0, T_n]\) represents continuous time over the inspiratory interval of duration \(T_n\).
The index \(n\) enumerates individual breaths in the dataset.

The signal \(x_n(t)\) is approximated by
\begin{equation}
x_n(t) \approx \sum_{k=1}^{M} f_{n,k}(t; \boldsymbol{\theta}_{n,k}),
\end{equation}
where \(M\) is the number of components used to represent the breath and \(k\) indexes the components within a single breath.
Each component captures a distinct substructure of the inspiratory waveform, such as a peak or ramp.

The function \(f_{n,k}(t; \boldsymbol{\theta}_{n,k})\) denotes the \(k\)-th component of the \(n\)-th breath and is parameterized by
\[
\boldsymbol{\theta}_{n,k} = (A_{n,k},\, t_{0,n,k},\, d_{n,k},\, \eta_{n,k}),
\]
where \(A_{n,k}\) is the component amplitude, \(t_{0,n,k}\) is the onset time, and \(d_{n,k}\) is the component duration.
The parameter vector \(\eta_{n,k}\) collects additional shape parameters specific to the chosen basis family, such as the standard deviation for a Gaussian basis or the shape parameters for a Beta basis.

For a given breath, the complete set of component parameters is denoted by
\[
\boldsymbol{\Theta}_n = \{\boldsymbol{\theta}_{n,1}, \dots, \boldsymbol{\theta}_{n,M}\}.
\]

\subsection{Basis Function Definitions}

A basis function defines a family of waveform shapes used to represent localized features within a signal.
Each family provides a parametric form that can be instantiated to model an individual event occurring over a finite time interval.
Each such instantiation is referred to as a \textit{component} and corresponds to one occurrence of the basis function in time.

To explicitly separate waveform shape from temporal support, each component is expressed as the product of a normalized basis function and a windowing function $w_{n,k}$.
The windowing function is defined as
\[
w_{n,k}(t) =
H\!\bigl(t - t_{0,n,k}\bigr)
-
H\!\bigl(t - (t_{0,n,k}+d_{n,k})\bigr),
\]
where \(H(\cdot)\) denotes the Heaviside step function, equal to 1 for nonnegative arguments and 0 otherwise.
This window restricts the component to be active only over the interval
\([t_{0,n,k},\, t_{0,n,k}+d_{n,k}]\).

Using this definition, the \(k\)-th component of the \(n\)-th signal is written as
\[
f_{n,k}(t; \boldsymbol{\theta}_{n,k}) =
A_{n,k}\,
\phi(t;\, t_{0,n,k},\, d_{n,k},\, \eta_{n,k})\,
w_{n,k}(t),
\]
where \(\phi(\cdot)\) denotes the normalized basis function that defines the waveform shape,
\(A_{n,k}\) is the component amplitude,
\(t_{0,n,k}\) is the onset time,
\(d_{n,k}\) is the duration,
and \(\eta_{n,k}\) represents any additional shape parameters specific to the basis family.

Three basis function families are considered in this work: \textbf{Gaussian}, \textbf{Half-Sine}, and \textbf{Beta}.

\subsubsection{Gaussian Basis Function}

\begin{figure}[!ht]
    \centering
    \includegraphics[width=\linewidth]{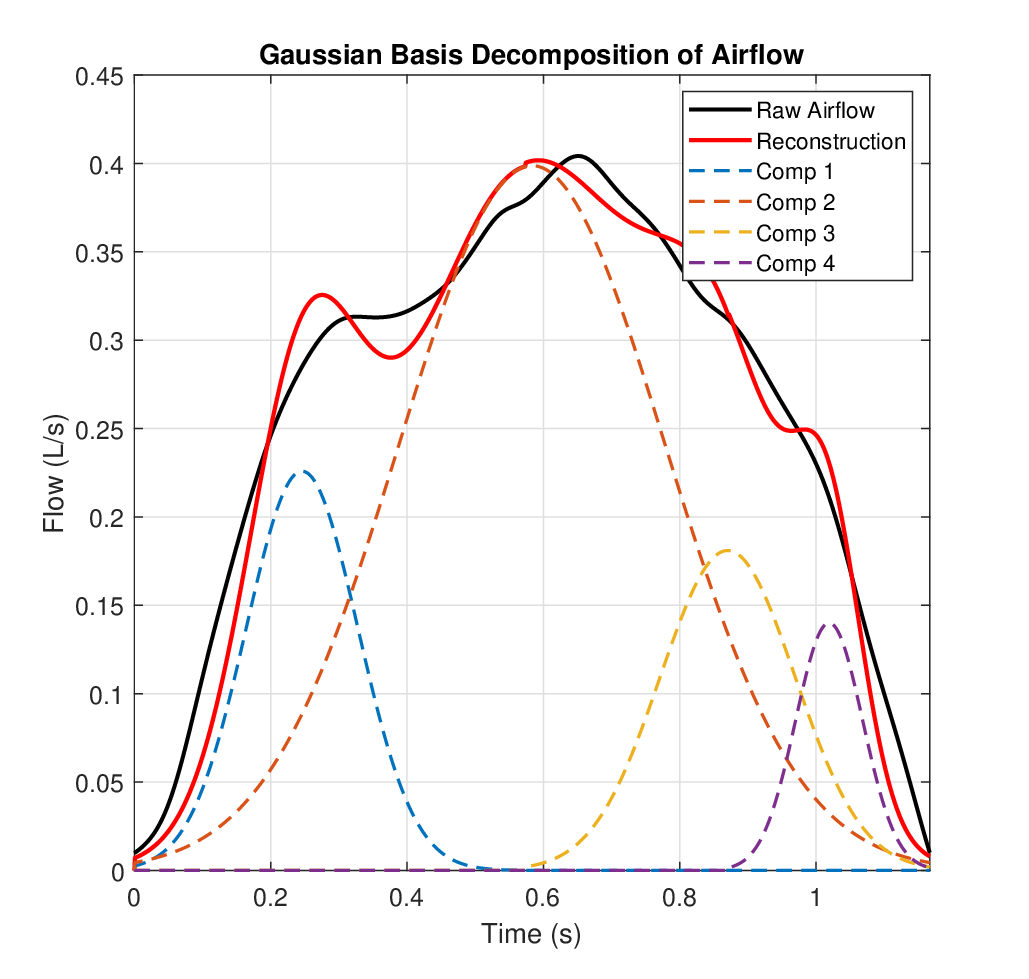}
    \caption{The raw airflow signal (black) is decomposed using the Gaussian basis function.  
    Each dashed curve represents an individual Gaussian component of the decomposition, defined by its amplitude \(A\), onset \(t_{0}\), and duration \(d = 6\sigma\).      The reconstructed signal (red) corresponds to the sum of all components.}
    \label{fig:gaussian_example}
\end{figure}

The Gaussian basis function is defined as
\[
\phi_{\mathrm{G}}(t;\, t_{0},\, d)
=
\exp\!\left[
-\frac{18\,(t - t_{0})^2}{d^2}
\right],
\]
where \(t_{0}\) denotes the center of the waveform and \(d\) denotes its effective duration.
The scaling factor is chosen such that \(d\) corresponds to the interval spanning approximately \(99.7\%\) of the Gaussian energy, equivalent to six standard deviations in the conventional parameterization.
A decomposition using the Gaussian basis function is illustrated in Figure~\ref{fig:gaussian_example}.

\subsubsection{Half-Sine Basis Function}

\begin{figure}[!ht]
    \centering
    \includegraphics[width=\linewidth]{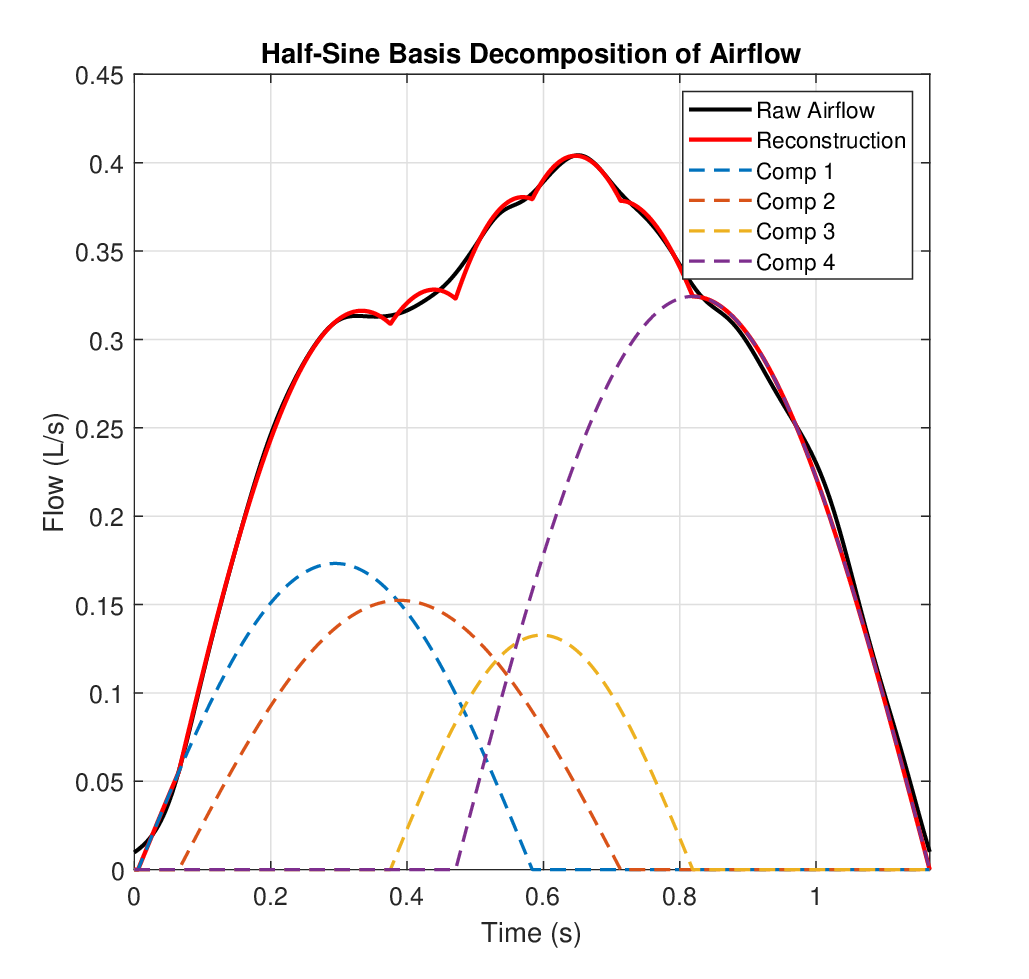}
    \caption{
    The raw airflow signal (black) is decomposed using the Half-Sine basis function.  
    Each dashed curve represents an individual Half-Sine component of the decomposition, defined by its amplitude \(A\), onset \(t_{0}\), and duration \(d\).  
    The reconstructed signal (red) corresponds to the sum of all components.
    }
    \label{fig:halfsine_example}
\end{figure}

The Half-Sine basis function is defined as
\[
\phi_{\mathrm{H}}(t;\, t_{0},\, d)
=
\sin\!\left(\frac{\pi (t - t_{0})}{d}\right),
\]
where \(t_{0}\) denotes the onset time and \(d\) denotes the duration.
A decomposition using the Half-Sine basis function is illustrated in Figure~\ref{fig:halfsine_example}.

\subsubsection{Beta Basis Function}

Let \(\tilde{x} = (t - t_{0}) / d\) denote the normalized time within the component support, with \(\tilde{x} \in [0,1]\).
For shape parameters \(\alpha > 1\) and \(\beta > 1\), define
\[
g(\tilde{x};\, \alpha,\, \beta)
= \tilde{x}^{\alpha-1}(1-\tilde{x})^{\beta-1}.
\]

The location of the maximum of \(g(\tilde{x};\, \alpha,\, \beta)\) is given by
\[
\tilde{x}_\ast
= \frac{\alpha-1}{\alpha+\beta-2}.
\]

The normalized Beta basis function is then defined as
\[
\phi_{\mathrm{B}}(t;\, t_{0},\, d,\, \alpha,\, \beta)
=
\frac{g(\tilde{x};\, \alpha,\, \beta)}
     {g(\tilde{x}_\ast;\, \alpha,\, \beta)},
\]
where \(\tilde{x} = (t - t_{0}) / d\).
This normalization ensures unit peak amplitude within the interval \([t_{0},\, t_{0}+d]\).

\begin{figure}[!ht]
    \centering
    \includegraphics[width=\linewidth]{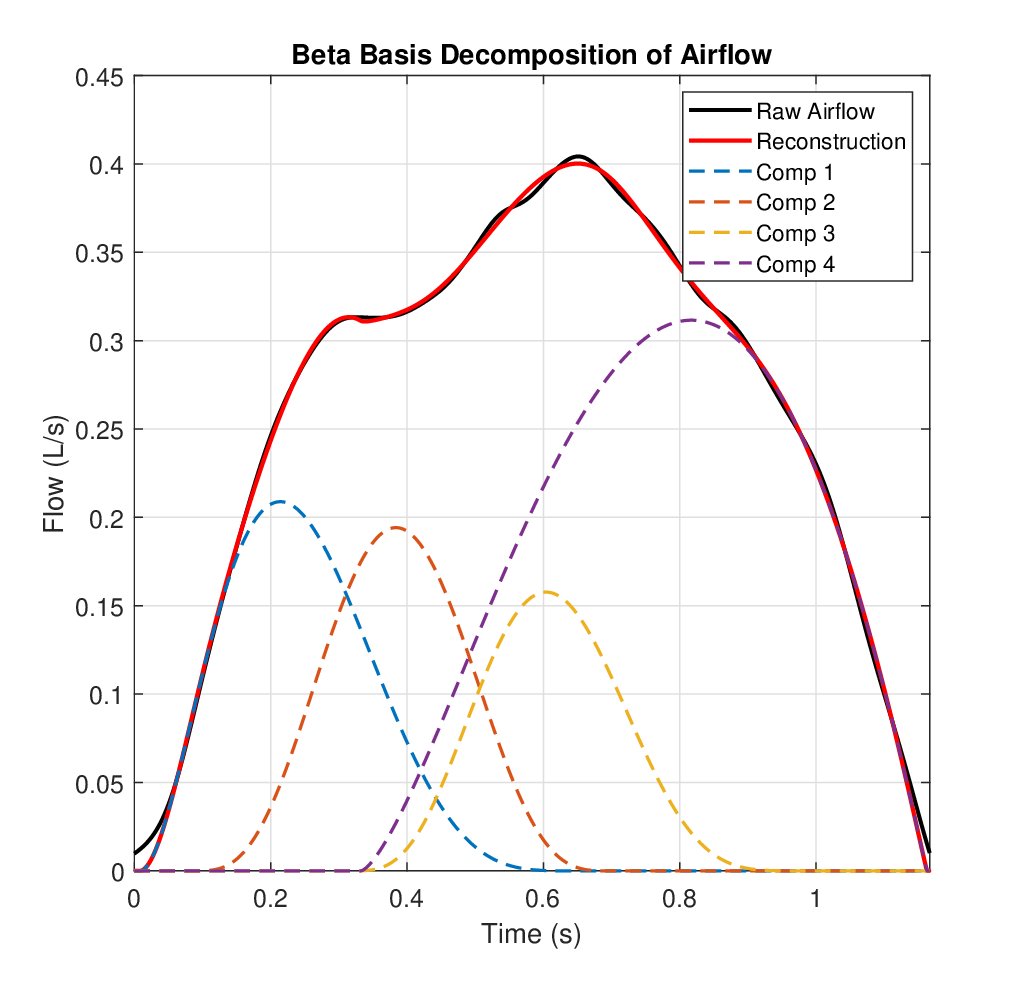}
    \caption{The raw airflow signal (black) is decomposed using the Beta basis function.  
    Each dashed curve represents an individual Beta component of the decomposition, defined by its amplitude \(A\), onset \(t_{0}\), duration \(d\), and shape parameters \(\alpha\) and \(\beta\).  
    The reconstructed signal (red) is the sum of all components.}
    \label{fig:beta_example}
\end{figure}

The parameters \(\alpha\) and \(\beta\) jointly control the waveform's asymmetry and peakedness.
When \(\alpha = \beta\), the basis function is symmetric, whereas unequal values introduce controlled skewness, producing either fast-rise/slow-decay or slow-rise/fast-decay profiles.
A decomposition using the Beta basis function is illustrated in Figure~\ref{fig:beta_example}.

\subsection{Optimization Framework}

For each breath $x_n(t)$, the set of component parameters
$\boldsymbol{\Theta}_n = \{\boldsymbol{\theta}_{n,1}, \dots, \boldsymbol{\theta}_{n,M}\}$
is estimated by minimizing the squared reconstruction error,
\begin{equation}
\begin{aligned}
\min_{\boldsymbol{\Theta}_n}
\quad
J_n
&=
\sum_t
\Bigg[
x_n(t)
-
\sum_{k=1}^{M}
f_{n,k}(t; \boldsymbol{\theta}_{n,k})
\Bigg]^2
\\
\text{subject to}
\quad
& 0 \le A_{n,k} \le A_{\max,n},
\\
& d_{\min} \le d_{n,k} \le T_n,
\\
& 0 \le t_{0,n,1} \le \delta,
\\
& 0 \le t_{0,n,k} \le T_n,
\qquad k = 2,\dots,M,
\end{aligned}
\end{equation}
where $A_{n,k}$, $d_{n,k}$, and $t_{0,n,k}$ denote the amplitude, duration, and onset time of the $k$th component,
respectively. The upper amplitude bound $A_{\max,n}$ is defined as the peak value of the airflow signal for breath $n$.
The minimum duration $d_{\min}=0.2$~s prevents non-physiological, impulse-like components, and $\delta = 0.001$ is a small
initial window that constrains the first component to activate near the onset of inspiration.

Each component is active only within its temporal support, as defined in the component's definition.
The constrained optimization problem is solved using MATLAB’s \texttt{fmincon} with the Interior-Point algorithm~\cite{Byrd1999, Byrd2000, Waltz2005}.

\subsubsection{Nonlinear Constraints}

To ensure components terminate within the breath window:

\[
t_{0,n,k} + d_{n,k} \le T_n,
\quad \forall\, k \in \{1,\dots,M\}.
\]

\subsubsection{Initial Conditions}

When no custom initialization is provided, component parameters are initialized using a simple, deterministic strategy that provides reasonable coverage of the inspiratory interval without biasing the solution toward a specific morphology. The amplitude of the \(k\)-th component is initialized as
\[
A_{n,k} = \frac{A_{\max,n}}{k},
\]
where \(A_{\max,n}\) denotes the peak amplitude of the inspiratory airflow signal. Component durations are initialized as decreasing fractions of the total inspiratory duration,
\[
d_{n,k} = \frac{T_n}{k},
\]
so that earlier components span broader portions of the breath while later components are progressively narrower.

The onset time of the first component is fixed at the start of inspiration (\(t_{0,n,1}=0\)). For subsequent components, onset times are initialized by evenly spacing them across the inspiratory interval,
\[
t_{0,n,k} = \frac{(k-1)\,T_n}{M+1}, \qquad k \geq 2,
\]
where \(M\) is the total number of components.

\subsubsection{Solver Settings}

Optimization tolerances (\texttt{FunctionTolerance} and \texttt{OptimalityTolerance}) are set to \(10^{-10}\),
with a maximum of \(10^4\) iterations. After convergence, components are sorted by onset time \(t_{0,n,k}\), and those with negligible amplitude (\(A_{n,k} < 0.001\)) are excluded. Reconstruction accuracy is reported using the mean squared error between \(x_n(t)\) and its fitted estimate.

\subsection{Experimental Protocol and Data Collection} 
\label{Sec:ExperimentalProtocol}

All study procedures were approved by the Institutional Review Board of the University of Florida (Protocol \# IRB202200508; IRB-01) on June 4, 2025, and were conducted in accordance with the Declaration of Helsinki. Prior to participation, individuals underwent pulmonary function testing, including forced vital capacity (FVC), forced expiratory volume in one second (FEV$_1$), maximal inspiratory pressure (MIP), and maximal expiratory pressure (MEP). At least three trials were performed for each measure, following the acceptability and reproducibility criteria of the American Thoracic Society and European Respiratory Society. The highest acceptable values were recorded, and participants had to achieve at least 75\% of the predicted values to qualify. 

Exclusion criteria included pregnancy, tobacco use, pulmonary disease (e.g., asthma, chronic obstructive pulmonary disease), or neurological disorders. A total of 25 healthy adults met the screening requirements and were enrolled in the study (6 females; age: $22.92 \pm 4.92$ years; BMI: $25.45 \pm 4.35$ kg/m$^2$).

Participants completed three trials of the Psychomotor Vigilance Task (PVT), each consisting of 70 cues and lasting approximately 10 minutes. The first trial served as a baseline, followed by either a low-load or high-load condition. The trial order was counterbalanced across participants. 

During the loaded trials, participants breathed through a linear airflow resistor (20 cmH\textsubscript{2}O/L/s) connected in series with a pressure-threshold valve set to 20\% of each subject’s maximal inspiratory pressure at the inspiratory port of the breathing circuit. In addition, a custom chest and abdominal restriction device applied 26 lbf in the lower-load condition and 52 lbf in the higher-load condition~\cite{ribeirorodriguesChest2025}. 

Airflow and mouth pressure were recorded using a Hans Rudolph pneumotachograph and differential pressure transducer. Two Millar pressure catheters were inserted nasally to measure esophageal and gastric pressures. Passive measures of chest wall and abdominal motion were obtained with g.RESPsensor belts (g.tec Medical Engineering). The upper belt was positioned on the chest just below the sternum to capture ribcage motion, and the lower belt was positioned on the abdomen above the navel. All signals were digitized at 1200 Hz using a g.HIamp biosignal amplifier.
\subsection{Derived Features}
\label{Sec:DerivedFeatures}

To evaluate the utility of the decomposition framework, a set of derived features was extracted from the component-level representation of each inspiratory breath. For every signal, the fitted parameters of the selected basis functions were first used to reconstruct the individual component waveforms. Both classical respiratory descriptors and decomposition-specific features were then computed from these reconstructed signals.

\subsubsection{Classical Respiratory Features}

The airflow signal, denoted by $\dot V(t)$, represents the time derivative of inspired volume and is expressed in liters per second (L/s). From this signal, several conventional respiratory features were derived to characterize the timing, magnitude, and shape of each inspiratory effort.

Inspiratory time, denoted by $T_i$, represents the duration of the inspiratory phase and is defined as the interval from the onset of positive airflow to its return to baseline. Tidal volume, denoted by $V_T$, quantifies the total inspired volume over this interval and is computed as
\[
V_{T} = \int_{0}^{T_{i}} \dot V(t)\, dt,
\]
where $\dot V(t)$ is the instantaneous airflow and $T_i$ defines the limits of integration.

Peak inspiratory flow, denoted by $\dot V_{\text{peak}}$, corresponds to the maximum instantaneous value of $\dot V(t)$ within the inspiratory phase. The symmetry index, denoted by $\mathrm{SI}$, characterizes the temporal balance of the inspiratory waveform and is defined as
\[
\mathrm{SI} = \frac{t_{\text{peak}}}{T_i},
\]
where $t_{\text{peak}}$ is the time from inspiratory onset to the occurrence of peak airflow.

The slope of rise to peak flow, denoted by $S_{\text{rise}}$, describes the rate at which airflow increases from onset to peak and is computed as
\[
S_{\text{rise}} = \frac{\dot V_{\text{peak}} - \dot V(0)}{t_{\text{peak}}},
\]
where $\dot V(0)$ is the airflow at inspiratory onset.

\subsubsection{Component-Level Features}

Each inspiratory breath was represented as the sum of $n_{\mathrm{comp}}$ time-localized basis functions. For each component $k$, three fundamental parameters were extracted directly from the fitted basis function.

The component amplitude, denoted by $A_k$, corresponds to the peak magnitude of the reconstructed component waveform and reflects its relative contribution to the total inspiratory airflow. The component duration, denoted by $d_k$, represents the temporal extent of the component and is defined by the support of the fitted basis function, indicating how long the subcomponent is active within the breath. The component onset time, denoted by $t_{0,k}$, specifies the initiation of the component relative to the start of inspiration.

To account for inter-breath variability in inspiratory duration, onset times were also normalized by the inspiratory time. The normalized onset time, denoted by $t^{*}_{0,k}$, is defined as
\[
t^{*}_{0,k} = \frac{t_{0,k}}{T_i},
\]
where $T_i$ is the inspiratory time of the corresponding breath. These parameters form the foundational representation of the decomposition and serve as inputs for higher-order derived features.

\subsubsection{Timing-Related Features}

Timing-related features were derived to quantify the temporal relationship between the raw airflow signal and its decomposed components. For each breath, the reconstructed component waveforms provided peak times denoted by $t^{(c)}_k$, where $k = 1,\dots,n_{\mathrm{comp}}$ indexes the components. In parallel, the raw airflow signal $\dot V(t)$ was analyzed to identify up to $n_{\mathrm{comp}}$ local maxima, with their corresponding peak times denoted by $t^{(r)}_j$, where $j = 1,\dots,n_{\mathrm{comp}}$. When fewer than four peaks were detected in the raw signal, the remaining entries were padded with zeros to preserve a consistent feature dimensionality across breaths.

The temporal offset between raw airflow peaks and component peaks was captured by the timing-difference matrix $\Delta T_{j,k}$, defined as
\[
\Delta T_{j,k} = t^{(r)}_j - t^{(c)}_k,
\]
where $t^{(r)}_j$ is the time of the $j$th raw airflow peak and $t^{(c)}_k$ is the time of the $k$th component peak.

From this matrix, the mean timing offset associated with each raw airflow peak was computed. The average timing difference, denoted by $\bar{\Delta} t^{(r)}_j$, is given by
\[
\bar{\Delta} t^{(r)}_j =
\frac{1}{n_{\mathrm{comp}}}
\sum_{k=1}^{n_{\mathrm{comp}}}
\Delta T_{j,k},
\quad
j = 1,\dots,n_{\mathrm{comp}}.
\]
This procedure yields up to four timing-offset features, $\bar{\Delta} t^{(r)}_1$ through $\bar{\Delta} t^{(r)}_4$. Values near zero indicate close temporal alignment between raw airflow peaks and component activations, whereas positive or negative values reflect delayed or advanced component activity relative to the raw signal.

Complementarily, the average timing offset from the component perspective was also computed as
% previously avgTimeDifComp_j
\[
\bar{\Delta} t^{(c)}_{k} = \frac{1}{n_{\mathrm{comp}}}
\sum_{j=1}^{n_{\mathrm{comp}}}
\Delta T_{j,k},
\quad 
k = 1, \dots, n_{\mathrm{comp}} .
\]
These features $(\bar{\Delta}t^{(c)}_{1} \text{–} \bar{\Delta} t^{(c)}_{4})$ represent the mean temporal difference of each component peak relative to all detected airflow peaks. Together, $\bar{\Delta}t^{(r)}$ and $\bar{\Delta}t^{(c)}$ provide a bidirectional characterization of timing alignment, capturing whether each component tends to precede or follow the corresponding airflow structure.
\subsubsection{Component–Component Timing Features}

To quantify the internal temporal structure of each breath, timing relationships among components were computed directly from their normalized onset times $t^{*}_{0,k}$. Pairwise differences between components' onsets were calculated as
% previously avgTimeDeltaCOmp_k
\[
\Delta t_{k,m} = t^{*}_{0,k} - t^{*}_{0,m},
\]
where \(k,m \in \{1, \dots, n_{\text{comp}}\}\).  
The average relative timing of each component was then obtained as
\[
\bar{\delta}t_{k} = 
\frac{1}{n_{\text{comp}} - 1} 
\sum_{\substack{m=1 \\ m \neq k}}^{n_{\text{comp}}} 
\Delta t_{k,m}.
\]
These features describe whether individual components tend to activate earlier or later relative to others within the same breath, providing a compact representation of phase coordination and sub-breath sequencing.

\subsubsection{Amplitude-Related Features}

Amplitude features were derived to quantify how closely the decomposition reproduced the magnitude of inspiratory airflow peaks. For each breath, the component waveforms yielded peak amplitudes $A^{(c)}_{k}$, and the raw airflow signal $\dot V(t)$ provided peak amplitudes $A^{(r)}_{j}$ for $j = 1, \dots, n_{\text{comp}}$.  
An amplitude-difference matrix was constructed as
\[
\Delta A_{j,k} = A^{(r)}_{j} - A^{(c)}_{k},
\]
which captures the offset in peak magnitude between each raw airflow peak ($A^{(r)}_{j}$) and each component peak ($A^{(c)}_{k}$).  When fewer than four peaks were detected in the raw signal, the remaining entries were padded with zeros up to the fourth position. These padded zeros were retained in all subsequent calculations to maintain dimensional consistency across breaths.
From this matrix, the mean amplitude difference across components for each raw peak was computed as
% previously avgAmpDifFlow_j
\[
\bar{\Delta}A^{(r)}_{j} = 
\frac{1}{n_{\text{comp}}} 
\sum_{k=1}^{n_{\text{comp}}} 
\Delta A_{j,k},
\quad 
j = 1, \dots, n_{\text{comp}} .
\]
Thus, up to four amplitude-difference features $(\bar{\Delta A}^{(r)}_{1} \text{–} \bar{\Delta A}^{(r)}_{4})$ were generated.  
Small values indicate that the decomposition reproduced the peak magnitude of the raw signal with high fidelity, whereas larger positive or negative values reflect systematic over- or underestimation of inspiratory peak strength.

Complementarily, the average amplitude difference from the component perspective was computed as
% previously avgAmpDifComp_j
\[
\bar{\Delta}A^{(c)}_{k} =
\frac{1}{n_{\text{comp}}}
\sum_{j=1}^{n_{\text{comp}}}
\Delta A_{j,k},
\quad 
k = 1, \dots, n_{\text{comp}} .
\]
These features $(\bar{\Delta A}^{(c)}_{1} \text{–} \bar{\Delta A}^{(c)}_{4})$ quantify the mean deviation of each component’s peak amplitude from all raw airflow peaks. Together, $\bar{\Delta}A^{(r)}$ and $\bar{\Delta}A^{(c)}$ provide a bidirectional measure of amplitude fidelity, indicating whether components tend to over- or underrepresent the magnitude of the airflow peaks.

\subsubsection{Component–Component Amplitude Features}

In addition to flow–component comparisons, amplitude coordination across components was quantified directly within the decomposition. Pairwise differences between component amplitudes were defined as
\[
\Delta A_{k,m} = A^{(c)}_{k} - A^{(c)}_{m},
\]
where \(k,m \in \{1, \dots, n_{\text{comp}}\}\).  
The mean amplitude difference for each component was then computed as
% previously avgAmpDeltaComp_k
\[
\bar{\delta} A_{k} =
\frac{1}{n_{\text{comp}} - 1}
\sum_{\substack{m=1 \\ m \neq k}}^{n_{\text{comp}}}
\Delta A_{k,m}.
\]
These features capture the distribution of relative magnitudes among subcomponents, describing how inspiratory effort is apportioned across the decomposed phases of the breath. When combined with the timing-delta metrics, they provide complementary information about the coordination of intensity and timing within the inspiratory pattern.
\begin{figure}
    \centering
    \includegraphics[width=\linewidth]{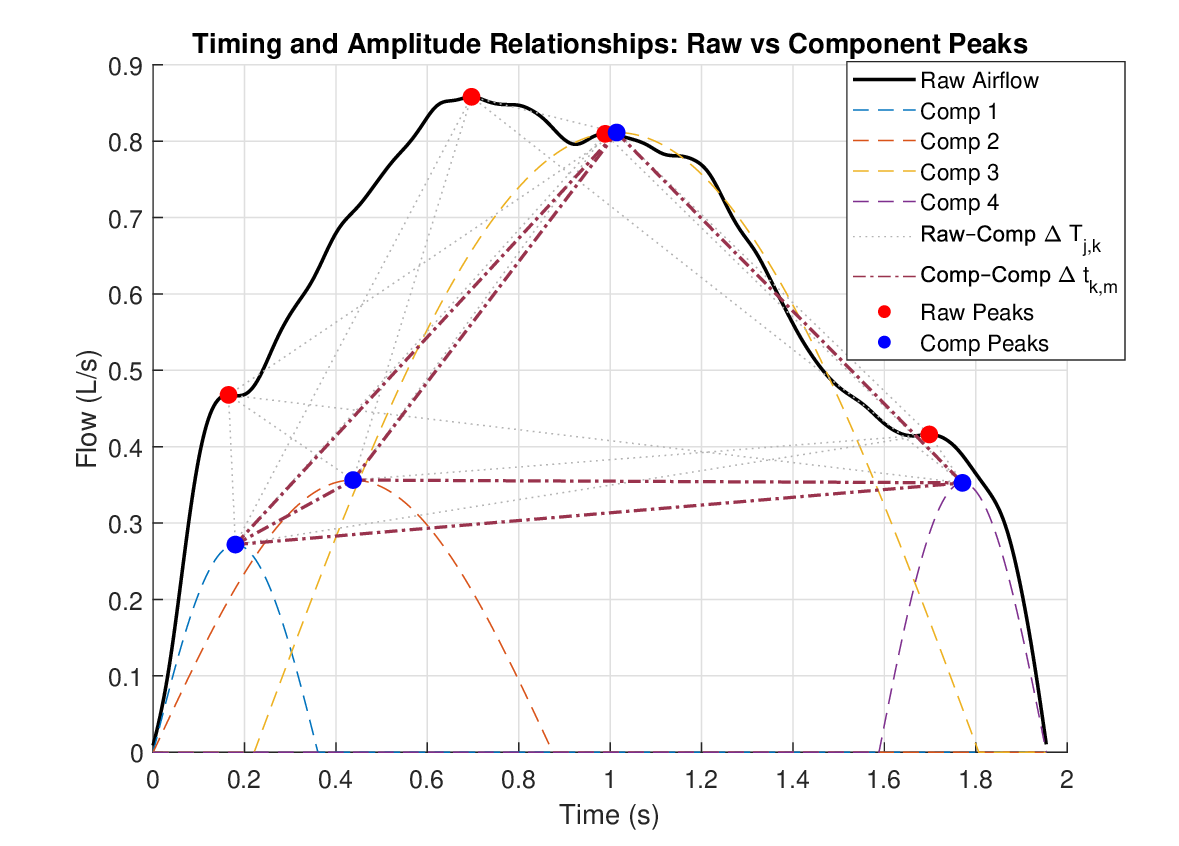}
\caption{Illustration of timing and amplitude relationships between the raw inspiratory airflow signal and its decomposed components. The raw airflow (black) is overlaid with reconstructed components (colored dashed lines). Red circles mark peaks in the raw signal, and blue circles mark component peaks. Gray dotted lines show the timing and amplitude differences between raw and component peaks ($\Delta T_{j,k}$), while red dashed--dot lines connect successive component peaks, indicating component--component timing differences ($\Delta t_{k,m}$).}
    \label{fig:example_timing}
\end{figure}
\subsubsection{Worked Example: Timing Features}

To illustrate the computation of timing-related features, consider one inspiratory cycle with four peaks detected in both the raw airflow signal and its reconstructed components.  
Although this example focuses on timing relationships (\(\Delta T_{j,k}\)), the same procedure applies to amplitude-based calculations (\(\Delta A_{j,k}\)), where peak magnitudes are substituted for peak times.  
Figure~\ref{fig:example_timing} visualizes this process, showing the alignment between raw airflow and component peaks as well as the timing differences used in the calculations.

\paragraph{Step 1: Peak detection}

Peaks were first detected in both the raw inspiratory airflow signal and the component waveforms to identify the key moments of maximal flow within the breath.  
Detected peak times in the raw airflow signal (in seconds):
\[
t^{(r)} = [0.1650,\; 0.6967,\; 0.9892,\; 1.6983].
\]

Detected peak times in the reconstructed components (in seconds):
\[
t^{(c)} = [0.1800,\; 0.4375,\; 1.0142,\; 1.7708].
\]
These times correspond to the red (raw) and blue (component) peaks shown in Figure~\ref{fig:example_timing}.

\paragraph{Step 2: Timing-difference matrix}

For each pair of raw and component peaks, the temporal offset is defined as
\[
\Delta T_{j,k} = t^{(r)}_{j} - t^{(c)}_{k},
\]
where \(j\) indexes raw peaks and \(k\) indexes component peaks.  
Numerically:
\[
\Delta T =
\begin{bmatrix}
-0.0150 & -0.2725 & -0.8492 & -1.6058\\
\;\;0.5167 & \;\;0.2592 & -0.3175 & -1.0742\\
\;\;0.8092 & \;\;0.5517 & -0.0250 & -0.7817\\
\;\;1.5183 & \;\;1.2608 & \;\;0.6842 & -0.0725
\end{bmatrix}\;\text{s}.
\]

Each element \(\Delta T_{j,k}\) represents how much earlier (negative) or later (positive) the raw peak occurs relative to the corresponding component peak.  
The gray dashed lines in Figure~\ref{fig:example_timing} indicate these pairwise temporal offsets.

\paragraph{Step 3: Mean offsets per raw peak}

The mean timing offset for each raw airflow peak is obtained by averaging across all component peaks:
\[
\bar{\Delta} t^{(r)}_{j} = 
\frac{1}{n_{\mathrm{comp}}} 
\sum_{k=1}^{n_{\mathrm{comp}}} 
\Delta T_{j,k}.
\]
For this example:
\[
\bar{\Delta} t^{(r)} = [-0.6856,\; -0.1540,\; 0.1385,\; 0.8477]\;\text{s}.
\]
In Figure~\ref{fig:example_timing}, these values correspond to the average horizontal separations between each red (raw) peak and all blue (component) peaks.

\paragraph{Step 4: Mean offsets per component peak}

Similarly, the mean timing offset for each component is computed by averaging across all raw peaks:
\[
\bar{\Delta} t^{(c)}_{k} = 
\frac{1}{n_{\mathrm{comp}}}
\sum_{j=1}^{n_{\mathrm{comp}}}
\Delta T_{j,k}.
\]
Resulting in:
\[
\bar{\Delta} t^{(c)} = [0.7073,\; 0.4498,\; -0.1269,\; -0.8835]\;\text{s}.
\]
These represent the average horizontal separations between each blue (component) peak and all red (raw) peaks.

\paragraph{Step 5: Component–component timing features}

Beyond the raw–component relationships, the internal sequencing of components is quantified using pairwise timing differences:
\[
\Delta t_{k,m} = t^{(c)}_{k} - t^{(c)}_{m},
\quad k,m \in \{1, \dots, n_{\text{comp}}\}.
\]
The purple dashed–dot lines in Figure~\ref{fig:example_timing} illustrate these component–component separations, which describe the temporal coordination among subcomponents of a single breath.  
For this example:
\[
\Delta t =
\begin{bmatrix}
0 & -0.2575 & -0.8342 & -1.5908\\
0.2575 & 0 & -0.5767 & -1.3333\\
0.8342 & 0.5767 & 0 & -0.7566\\
1.5908 & 1.3333 & 0.7566 & 0
\end{bmatrix}\;\text{s}.
\]
The mean relative activation time for each component is then
\[
\bar{\delta}t_{k} = 
\frac{1}{n_{\text{comp}} - 1} 
\sum_{\substack{m=1 \\ m \neq k}}^{n_{\text{comp}}} 
\Delta t_{k,m},
\]
yielding
\[
\bar{\delta}t = [-0.8941,\; -0.5508,\; 0.3892,\; 1.0557]\;\text{s}.
\]

\subsection{Classification Task Setup}

The classification analysis was conducted to evaluate whether breaths could be distinguished as fatigued (F) or non-fatigued (NF) under high inspiratory load, where fatigue reflects cognitive-respiratory competition rather than respiratory muscle fatigue. Specifically, we investigated whether respiratory airflow patterns reflected the diversion of cognitive resources toward breathing control, as indexed by concurrent decline in vigilance task performance. The underlying hypothesis is that when automatic respiratory compensation to mechanical load is insufficient, breathing may require increased cognitive resources and attentional control, potentially competing with concurrent cognitive task performance and leading to cognitive fatigue. Data were drawn from the high-load trials. This experiment served as a proof-of-concept application of the proposed decomposition framework, providing a practical test of whether the derived features capture breathing pattern changes associated with cognitive-respiratory competition beyond conventional respiratory measures.

Fatigue classification was determined at the subject level based on PVT performance during loaded breathing. For each subject, response times were filtered to remove anticipatory responses ($<$100 ms) and outliers (below the 5th or above the 95th percentile), then fit with a linear regression against elapsed time. The slope of this fit, expressed in milliseconds per minute, served as an index of vigilance decline during loaded breathing—a marker of cognitive fatigue. Subjects with positive slopes, indicating progressively slower reaction times, were interpreted as experiencing increased cognitive demand from breathing that interfered with task performance, suggesting incomplete automatic compensation to the respiratory load and resulting cognitive fatigue from dual-task competition. A total of 10 participants out of 25 met this criterion.

Within these fatigued subjects, individual breaths were labeled according to their temporal occurrence during the trial: breaths from the early phase (10–20\% of total duration) were defined as non-fatigued (NF), representing the initial state when compensation strategies were being established and cognitive-respiratory competition was minimal, whereas breaths from the late phase ($>$90\%) were defined as fatigued (F), representing the state when cognitive resources were increasingly diverted to maintain breathing and cognitive fatigue had developed. Initial breaths in the first 0–10\% of the trial were excluded to account for adjustment to the respiratory load. The final dataset contained a comparable number of NF (192) and F (183) samples. No additional resampling, weighting, or balancing procedures were applied. Throughout this analysis, the terms ``fatigued'' and ``non-fatigued'' refer to states of cognitive fatigue arising from cognitive-respiratory competition, not to respiratory muscle fatigue or diminished respiratory pump capacity.

Two feature pools were defined for the classification experiments. The \textit{classical} pool included conventional respiratory features ($T_{i}$, $V_{T}$, $\dot V_{\text{peak}}$, $\text{SI}$, and $S_{\text{rise}}$) representing standard descriptors of inspiratory timing, volume, and raw airflow shape. The \textit{composite} pool included all decomposition-derived metrics in addition to these classical features, encompassing amplitude and timing-based descriptors calculated from the reconstructed component waveforms.

Classification trees were trained using the \texttt{fitctree} function (MATLAB R2024b, MathWorks, Natick, MA) with interaction-curvature predictor selection and a minimum leaf size of 30. For each feature pool, all unique combinations of two, three, and four features were evaluated to systematically assess the influence of feature type and dimensionality on discriminative performance. A leave-one-subject-out (LOSO) cross-validation scheme was applied, training models on $N-1$ participants and testing on the held-out participant. Performance was quantified using the Matthews correlation coefficient (MCC) and $F_{1}$ score.
\section{Results}

\subsection{RQ1 - Can inspiratory airflow be reconstructed at the single-breath level using a small number of time-localized, interpretable components?}

To address this question, each inspiratory breath was decomposed using the Half-Sine, Gaussian, and Beta basis functions across one to four components. For each model and condition, the reconstruction accuracy was quantified using the mean squared error (MSE) between the original and reconstructed airflow signals. Results were aggregated across all 8276 breaths (4077 under load).

\begin{figure*}[!ht]
\centering
\begin{subfigure}{0.48\textwidth}
    \includegraphics[width=\linewidth]{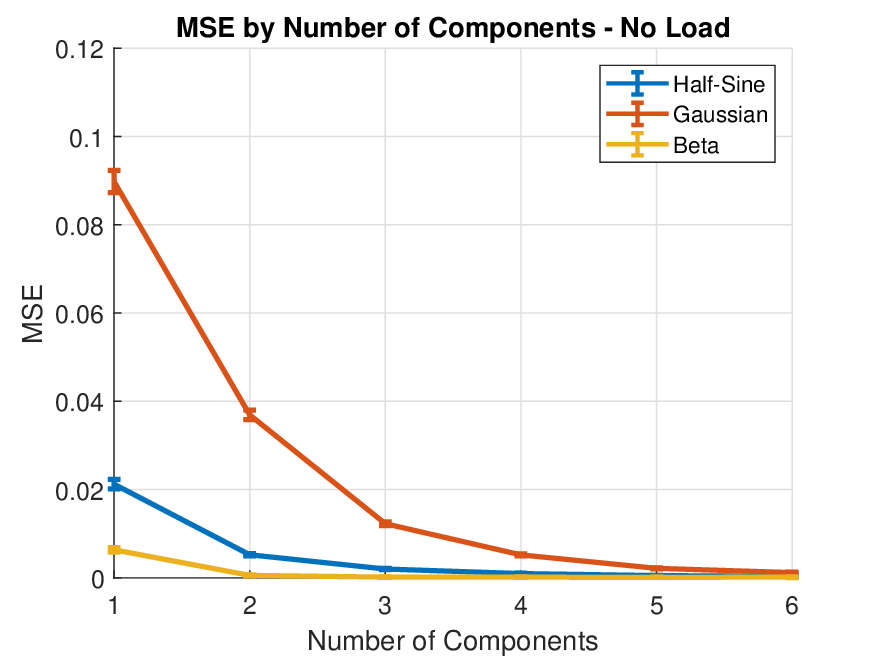}
    \caption{No-load breaths}
    \label{fig:mse_noload}
\end{subfigure}
\hfill
\begin{subfigure}{0.48\textwidth}
    \includegraphics[width=\linewidth]{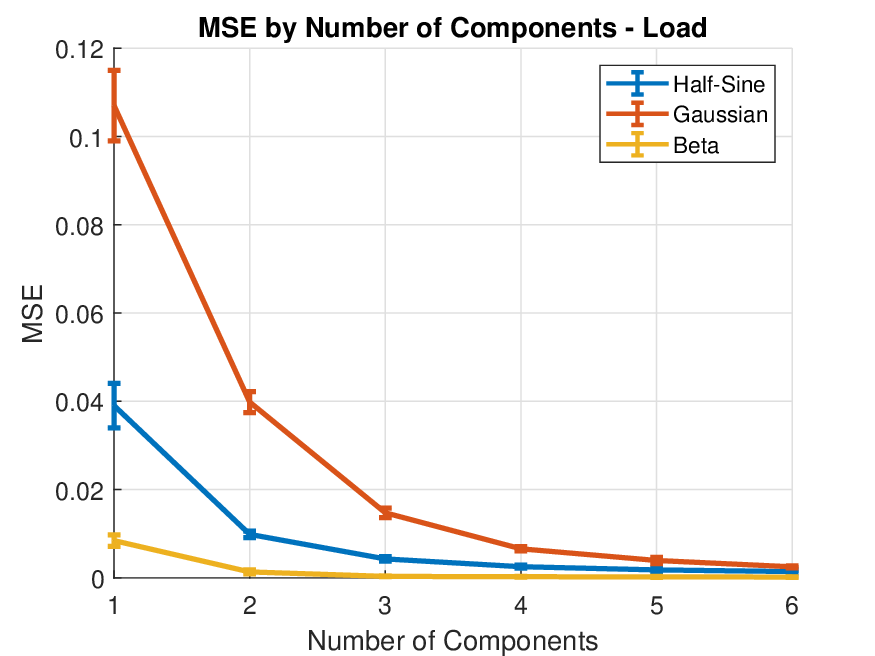}
    \caption{Loaded breaths}
    \label{fig:mse_load}
\end{subfigure}
\caption{Mean squared error by number of components for the Half-Sine, Gaussian, and Beta models. Error bars indicate 95\% confidence intervals.}
\label{fig:mse_comparison}
\end{figure*}

Across all basis function families, MSE decreased monotonically with increasing component count (Fig.~\ref{fig:mse_comparison}). Tables~\ref{tab:mse_noload} and~\ref{tab:mse_load} summarize the MSE values and 95\% confidence intervals for the no-load and load conditions, respectively. 

\begin{table}[!ht]
\centering
\resizebox{\columnwidth}{!}{%
\begin{tabular}{@{}lccc@{}}
\toprule
\multicolumn{4}{c}{\textbf{MSE — No Load Condition}} \\ \midrule
 & \textbf{Half-Sine} & \textbf{Gaussian} & \textbf{Beta} \\ \midrule
1 Component  & $0.0212 \pm 0.0011$ & $0.0898 \pm 0.0025$ & $0.0064 \pm 0.0006$ \\
2 Components & $0.0052 \pm 0.0003$ & $0.0369 \pm 0.0011$ & $0.0006 \pm 0.0001$ \\
3 Components & $0.0020 \pm 0.0001$ & $0.0123 \pm 0.0005$ & $0.0002 \pm 0.0001$ \\
4 Components & $0.0010 \pm 0.0001$ & $0.0052 \pm 0.0002$ & $0.0001 \pm 0.0001$ \\ 
5 Components & $0.0005 \pm 0.0001$ & $0.0022 \pm 0.0002$ & $0.0001 \pm 0.0001$ \\
6 Components & $0.0003 \pm 0.0001$ & $0.0012 \pm 0.0001$ & $0.0002 \pm 0.0001$ \\ 
\bottomrule
\end{tabular}%
}
\caption{Reconstruction MSE across basis function families under the no-load condition. Values are reported as mean $\pm$ 95\% confidence interval half-width.}
\label{tab:mse_noload}
\end{table}

\begin{table}[!ht]
\centering
\resizebox{\columnwidth}{!}{%
\begin{tabular}{@{}lccc@{}}
\toprule
\multicolumn{4}{c}{\textbf{MSE — Load Condition}} \\ \midrule
 & \textbf{Half-Sine} & \textbf{Gaussian} & \textbf{Beta} \\ \midrule
1 Component  & $0.0390 \pm 0.0051$ & $0.1070 \pm 0.0080$ & $0.0084 \pm 0.0013$ \\
2 Components & $0.0099 \pm 0.0008$ & $0.0398 \pm 0.0024$ & $0.0013 \pm 0.0005$ \\
3 Components & $0.0043 \pm 0.0005$ & $0.0147 \pm 0.0011$ & $0.0003 \pm 0.0001$ \\
4 Components & $0.0025 \pm 0.0004$ & $0.0066 \pm 0.0005$ & $0.0003 \pm 0.0001$ \\ 
5 Components & $0.0018 \pm 0.0004$ & $0.0039 \pm 0.0007$ & $0.0002 \pm 0.0002$ \\
6 Components & $0.0014 \pm 0.0003$ & $0.0025 \pm 0.0002$ & $0.0002 \pm 0.0001$ \\ 
\bottomrule
\end{tabular}%
}
\caption{Reconstruction MSE across basis function families under the load condition. Values are reported as mean $\pm$ 95\% confidence interval half-width.}
\label{tab:mse_load}
\end{table}

Friedman tests performed separately for each component order showed a significant main effect of basis function on MSE (all $p < 0.001$). Post-hoc comparisons confirmed that the Half-Sine, Gaussian, and Beta models differed significantly from one another at all component counts (all $p < 0.001$). 

\subsection{RQ2 - Does the proposed parametric decomposition yield precise and repeatable component parameters under moderate noise?} 

A total of 8276 breaths were available. To reduce runtime, a random sample of 1600 breaths was selected without replacement. Each selected breath was decomposed using three models with orders ranging from 1 to 4 components: Half-Sine, Gaussian, and Beta. For Half-Sine and Gaussian, the reported parameters are amplitude, duration, and onset time. For the Beta model the reported parameters are amplitude, duration, onset time, and the Beta shape parameters \(\alpha\) and \(\beta\).
For each breath and model order, an original fit yielded the parameter vector \texttt{original\_params}. To assess robustness to moderate noise, zero-mean Gaussian noise with standard deviation equal to 5\% of the signal standard deviation (average SNR = 30 dB) was added to the raw airflow, and the full decomposition was repeated 50 times for each breath.

Across the noisy runs, the mean and standard deviation of each parameter were computed for each component index. For summarization across breaths, the mean of the original parameters and the mean of the noisy means were computed for each component index. Across the 100 analyzed breaths, decomposition using the Half-Sine, Gaussian, and Beta models showed close agreement between the original and noisy parameter estimates under white-noise perturbation. For the Half-Sine model (Table~\ref{tab:halfsine_params}), amplitude, duration, and onset time remained consistent across one to four components, with mean noisy values closely matching the originals and standard deviations generally below 0.19. The Gaussian model (Table~\ref{tab:gaussian_params}) yielded more stable estimates, with amplitude, duration, and onset time showing minimal variability, with standard deviations typically below 0.04 and no systematic deviations across orders. The Beta model (Table~\ref{tab:beta_params_full}) yielded stable amplitude, duration, and onset time estimates with standard deviations below 0.16, while the additional shape parameters \(\alpha\) and \(\beta\) displayed greater variability, with standard deviations ranging from 0.4 to 1.6 at higher component orders.

\begin{table}[!ht]
\centering
\resizebox{\columnwidth}{!}{%
\begin{tabular}{@{}lcccc@{}}
\toprule
\multicolumn{5}{c}{\textbf{Half-Sine Model}} \\ \midrule
 & \multicolumn{1}{l}{Component 1} & \multicolumn{1}{l}{Component 2} & \multicolumn{1}{l}{Component 3} & \multicolumn{1}{l}{Component 4} \\ \midrule
\multicolumn{5}{c}{\textbf{Amplitude}} \\ \midrule
1 Component  & 0.72 (0.73 $\pm$ 0.00) & - & - & - \\
2 Components & 0.57 (0.59 $\pm$ 0.03) & 0.51 (0.54 $\pm$ 0.04) & - & - \\
3 Components & 0.47 (0.49 $\pm$ 0.08) & 0.48 (0.51 $\pm$ 0.07) & 0.47 (0.47 $\pm$ 0.07) & - \\
4 Components & 0.45 (0.46 $\pm$ 0.11) & 0.39 (0.38 $\pm$ 0.11) & 0.34 (0.35 $\pm$ 0.10) & 0.45 (0.45 $\pm$ 0.08) \\ \midrule
\multicolumn{5}{c}{\textbf{Duration}} \\ \midrule
1 Component  & 1.50 (1.50 $\pm$ 0.00) & - & - & - \\
2 Components & 1.12 (1.08 $\pm$ 0.06) & 1.03 (0.96 $\pm$ 0.06) & - & - \\
3 Components & 0.88 (0.85 $\pm$ 0.13) & 0.95 (0.92 $\pm$ 0.14) & 0.64 (0.62 $\pm$ 0.08) & - \\
4 Components & 0.79 (0.80 $\pm$ 0.18) & 0.87 (0.87 $\pm$ 0.19) & 0.63 (0.64 $\pm$ 0.16) & 0.54 (0.54 $\pm$ 0.10) \\ \midrule
\multicolumn{5}{c}{\textbf{Onset Time}} \\ \midrule
1 Component  & 0.01 (0.01 $\pm$ 0.00) & - & - & - \\
2 Components & 0.01 (0.01 $\pm$ 0.00) & 0.49 (0.57 $\pm$ 0.05) & - & - \\
3 Components & 0.01 (0.01 $\pm$ 0.00) & 0.28 (0.31 $\pm$ 0.09) & 0.91 (0.93 $\pm$ 0.08) & - \\
4 Components & 0.01 (0.01 $\pm$ 0.00) & 0.22 (0.22 $\pm$ 0.10) & 0.62 (0.62 $\pm$ 0.12) & 0.99 (0.99 $\pm$ 0.09) \\ 
\bottomrule
\end{tabular}%
}
\caption{Mean, mean of noisy estimates, and standard deviation of amplitude, duration, and onset time for each component across decomposition orders (1–4) using the Half-Sine model. Values are reported as: mean(original) (mean(noisy) $\pm$ SD(noisy)).}
\label{tab:halfsine_params}
\end{table}

\begin{table}[!ht]
\centering
\resizebox{\columnwidth}{!}{%
\begin{tabular}{@{}lcccc@{}}
\toprule
\multicolumn{5}{c}{\textbf{Gaussian Model}} \\ \midrule
 & \multicolumn{1}{l}{Component 1} & \multicolumn{1}{l}{Component 2} & \multicolumn{1}{l}{Component 3} & \multicolumn{1}{l}{Component 4} \\ \midrule
\multicolumn{5}{c}{\textbf{Amplitude}} \\ \midrule
1 Component  & 0.73 (0.76 $\pm$ 0.01) & - & - & - \\
2 Components & 0.70 (0.72 $\pm$ 0.01) & 0.68 (0.70 $\pm$ 0.01) & - & - \\
3 Components & 0.60 (0.59 $\pm$ 0.02) & 0.69 (0.70 $\pm$ 0.01) & 0.54 (0.54 $\pm$ 0.01) & - \\
4 Components & 0.46 (0.47 $\pm$ 0.03) & 0.64 (0.64 $\pm$ 0.03) & 0.56 (0.56 $\pm$ 0.03) & 0.43 (0.43 $\pm$ 0.02) \\ \midrule
\multicolumn{5}{c}{\textbf{Duration}} \\ \midrule
1 Component  & 1.51 (1.51 $\pm$ 0.00) & - & - & - \\
2 Components & 1.07 (1.06 $\pm$ 0.01) & 1.13 (1.12 $\pm$ 0.01) & - & - \\
3 Components & 0.77 (0.76 $\pm$ 0.02) & 1.26 (1.26 $\pm$ 0.03) & 0.70 (0.70 $\pm$ 0.02) & - \\
4 Components & 0.60 (0.60 $\pm$ 0.04) & 1.09 (1.10 $\pm$ 0.06) & 1.00 (0.99 $\pm$ 0.05) & 0.49 (0.49 $\pm$ 0.03) \\ \midrule
\multicolumn{5}{c}{\textbf{Onset Time}} \\ \midrule
1 Component  & 0.00 (0.01 $\pm$ 0.00) & - & - & - \\
2 Components & 0.00 (0.00 $\pm$ 0.00) & 0.43 (0.44 $\pm$ 0.01) & - & - \\
3 Components & 0.00 (0.00 $\pm$ 0.00) & 0.16 (0.16 $\pm$ 0.01) & 0.86 (0.86 $\pm$ 0.02) & - \\
4 Components & 0.00 (0.00 $\pm$ 0.00) & 0.06 (0.06 $\pm$ 0.01) & 0.50 (0.50 $\pm$ 0.04) & 1.07 (1.07 $\pm$ 0.02) \\ 
\bottomrule
\end{tabular}%
}
\caption{Mean, mean of noisy estimates, and standard deviation of amplitude, duration, and onset time for each component across decomposition orders (1–4) using the Gaussian model. Values are reported as: mean(original) (mean(noisy) $\pm$ SD(noisy)).}
\label{tab:gaussian_params}
\end{table}

\begin{table}[!ht]
\centering
\resizebox{\columnwidth}{!}{%
\begin{tabular}{@{}lcccc@{}}
\toprule
\multicolumn{5}{c}{\textbf{Beta Model}} \\ \midrule
 & \multicolumn{1}{l}{Component 1} & \multicolumn{1}{l}{Component 2} & \multicolumn{1}{l}{Component 3} & \multicolumn{1}{l}{Component 4} \\ \midrule
\multicolumn{5}{c}{\textbf{Amplitude}} \\ \midrule
1 Component  & 0.70 (0.71 $\pm$ 0.01) & - & - & - \\
2 Components & 0.53 (0.54 $\pm$ 0.06) & 0.50 (0.50 $\pm$ 0.06) & - & - \\
3 Components & 0.48 (0.48 $\pm$ 0.08) & 0.44 (0.45 $\pm$ 0.10) & 0.46 (0.46 $\pm$ 0.08) & - \\
4 Components & 0.45 (0.46 $\pm$ 0.08) & 0.41 (0.41 $\pm$ 0.11) & 0.40 (0.40 $\pm$ 0.12) & 0.42 (0.42 $\pm$ 0.10) \\ \midrule
\multicolumn{5}{c}{\textbf{Duration}} \\ \midrule
1 Component  & 1.52 (1.52 $\pm$ 0.02) & - & - & - \\
2 Components & 1.41 (1.40 $\pm$ 0.06) & 1.26 (1.26 $\pm$ 0.08) & - & - \\
3 Components & 1.10 (1.09 $\pm$ 0.16) & 1.15 (1.14 $\pm$ 0.15) & 1.00 (0.99 $\pm$ 0.13) & - \\
4 Components & 0.93 (0.92 $\pm$ 0.14) & 0.94 (0.92 $\pm$ 0.15) & 0.96 (0.95 $\pm$ 0.12) & 0.87 (0.87 $\pm$ 0.12) \\ \midrule
\multicolumn{5}{c}{\textbf{Onset Time}} \\ \midrule
1 Component  & 0.01 (0.01 $\pm$ 0.00) & - & - & - \\
2 Components & 0.00 (0.00 $\pm$ 0.00) & 0.09 (0.09 $\pm$ 0.03) & - & - \\
3 Components & 0.00 (0.00 $\pm$ 0.00) & 0.14 (0.14 $\pm$ 0.05) & 0.49 (0.50 $\pm$ 0.11) & - \\
4 Components & 0.00 (0.00 $\pm$ 0.00) & 0.14 (0.15 $\pm$ 0.05) & 0.39 (0.40 $\pm$ 0.08) & 0.65 (0.65 $\pm$ 0.11) \\ \midrule
\multicolumn{5}{c}{\textbf{Alpha}} \\ \midrule
1 Component  & 1.70 (1.72 $\pm$ 0.04) & - & - & - \\
2 Components & 3.32 (3.36 $\pm$ 0.65) & 2.81 (2.85 $\pm$ 0.47) & - & - \\
3 Components & 3.09 (3.11 $\pm$ 0.94) & 4.01 (4.11 $\pm$ 1.51) & 3.04 (3.08 $\pm$ 0.83) & - \\
4 Components & 3.44 (3.46 $\pm$ 1.05) & 3.69 (3.81 $\pm$ 1.37) & 4.18 (4.22 $\pm$ 1.56) & 4.21 (4.24 $\pm$ 1.50) \\ \midrule
\multicolumn{5}{c}{\textbf{Beta}} \\ \midrule
1 Component  & 1.66 (1.68 $\pm$ 0.04) & - & - & - \\
2 Components & 2.56 (2.56 $\pm$ 0.39) & 3.11 (3.11 $\pm$ 0.48) & - & - \\
3 Components & 4.14 (4.22 $\pm$ 1.19) & 3.72 (3.78 $\pm$ 1.30) & 2.38 (2.37 $\pm$ 0.70) & - \\
4 Components & 4.63 (4.65 $\pm$ 1.31) & 4.48 (4.57 $\pm$ 1.60) & 3.72 (3.74 $\pm$ 1.43) & 2.48 (2.53 $\pm$ 0.94) \\ 
\bottomrule
\end{tabular}%
}
\caption{Mean, mean of noisy estimates, and standard deviation of amplitude, duration, onset time, and Beta distribution shape parameters (\(\alpha\), \(\beta\)) for each component across decomposition orders (1–4) using the Beta model. Values are reported as: mean(original) (mean(noisy) $\pm$ SD(noisy)).}
\label{tab:beta_params_full}
\end{table}

\subsection{RQ3 - Do sub-breath features derived from the proposed representation capture meaningful adaptations in breathing dynamics that are not accessible through conventional breath-level measures?}

Classification models were trained to distinguish fatigued from non-fatigued breaths using two different feature pools: (1) the classical set including $T_i$, $V_T$, $\dot V_{\text{peak}}$, SI and $S_{\text{rise}}$; and (2) a composite set including all decomposition-derived features in addition to the classical ones.

Each model used leave-one-subject-out cross-validation and was trained using all unique combinations of two, three, and four features. Performance was evaluated with the MCC and $F_1$ score.

For the classical feature pool, the best two-feature model ($T_i$, $\dot V_{\text{peak}}$) achieved MCC = 0.419 and $F_1$ = 0.686. The best three-feature model ($T_i$, $\dot V_{\text{peak}}$, SI) achieved identical performance (MCC = 0.419, $F_1$ = 0.686). The best four-feature model ($T_i$, $\dot V_{\text{peak}}$, SI, $S_{\text{rise}}$) achieved MCC = 0.414 and $F_1$ = 0.680.

\begin{figure*}[!ht]
\centering
\setlength{\tabcolsep}{2pt}
% --- Classical Models ---
\begin{subfigure}[t]{0.32\textwidth}
    \centering
    \includegraphics[width=\linewidth]{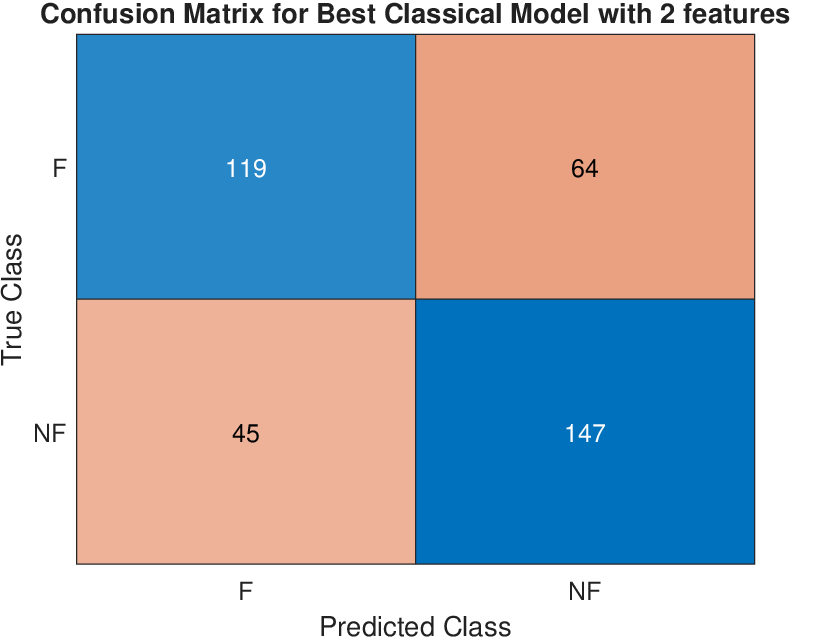}
    \caption{Classical, 2 features: $T_i$, $\dot V_{\text{peak}}$}
\end{subfigure}
\hfill
\begin{subfigure}[t]{0.32\textwidth}
    \centering
    \includegraphics[width=\linewidth]{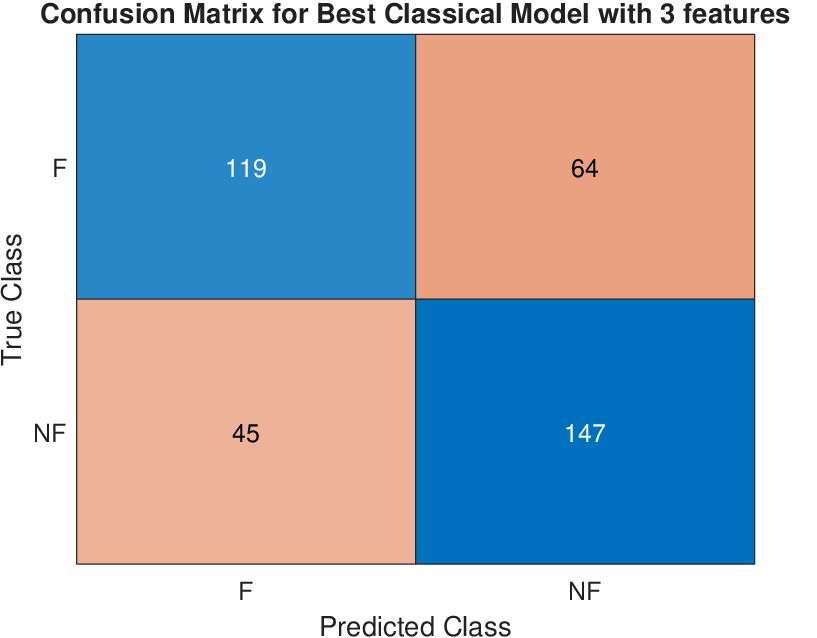}
    \caption{Classical, 3 features: $T_i$, $\dot V_{\text{peak}}$, SI}
\end{subfigure}
\hfill
\begin{subfigure}[t]{0.32\textwidth}
    \centering
    \includegraphics[width=\linewidth]{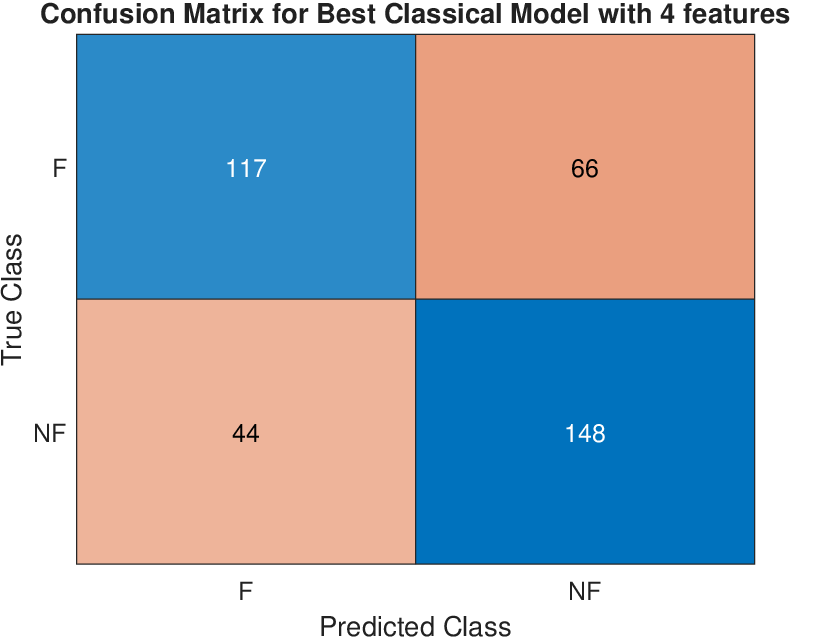}
    \caption{Classical, 4 features: $T_i$, $\dot V_{\text{peak}}$, SI, $S_{\text{rise}}$}
\end{subfigure}

\vspace{4pt}

% --- Composite Models ---
\begin{subfigure}[t]{0.32\textwidth}
    \centering
    \includegraphics[width=\linewidth]{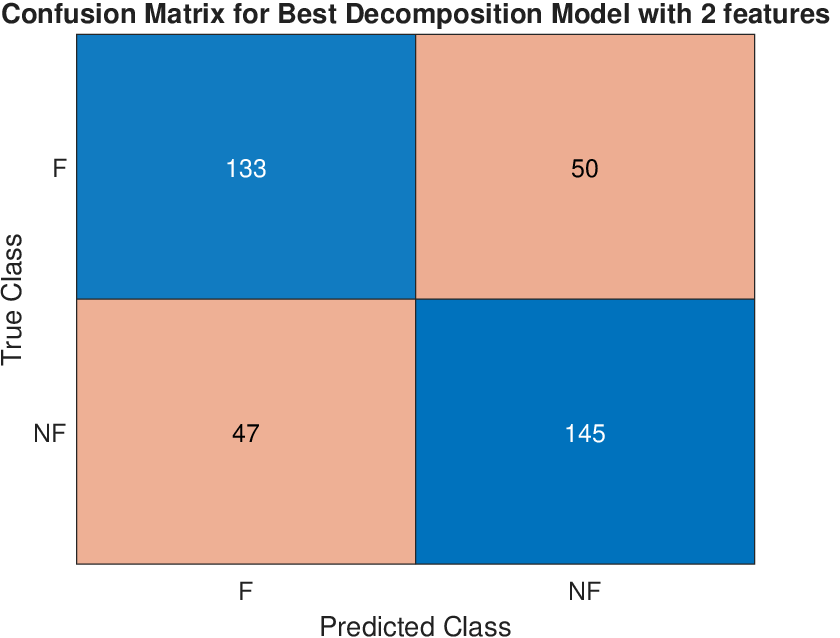}
    \caption{Composite, 2 features: $\bar{\Delta}t^{(r)}_{1}$, $\bar{\delta}t_2$}
\end{subfigure}
\hfill
\begin{subfigure}[t]{0.32\textwidth}
    \centering
    \includegraphics[width=\linewidth]{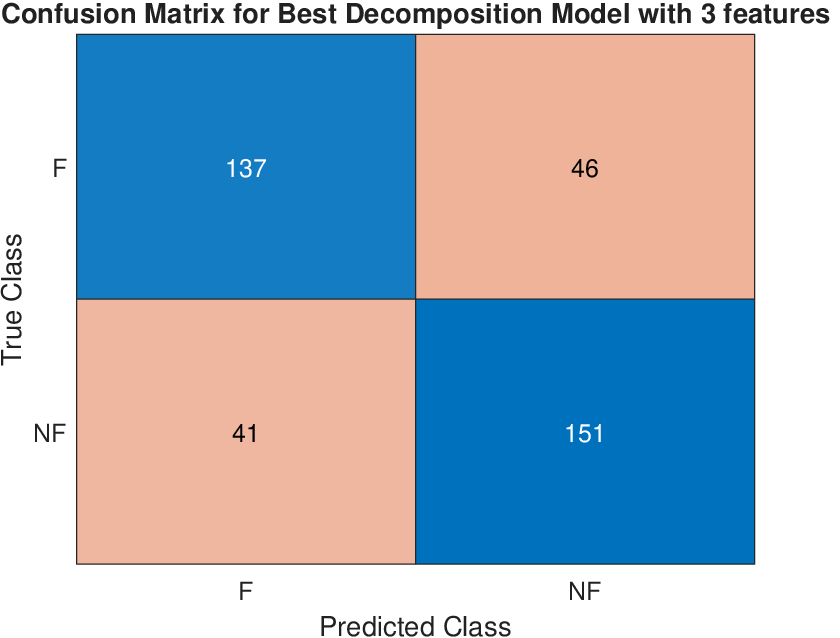}
    \caption{Composite, 3 features: $\bar{\Delta} t^{(c)}_1$,$\bar{\Delta}t^{(r)}_{1}$, $\bar{\Delta}A^{(r)}_4$}
\end{subfigure}
\hfill
\begin{subfigure}[t]{0.32\textwidth}
    \centering
    \includegraphics[width=\linewidth]{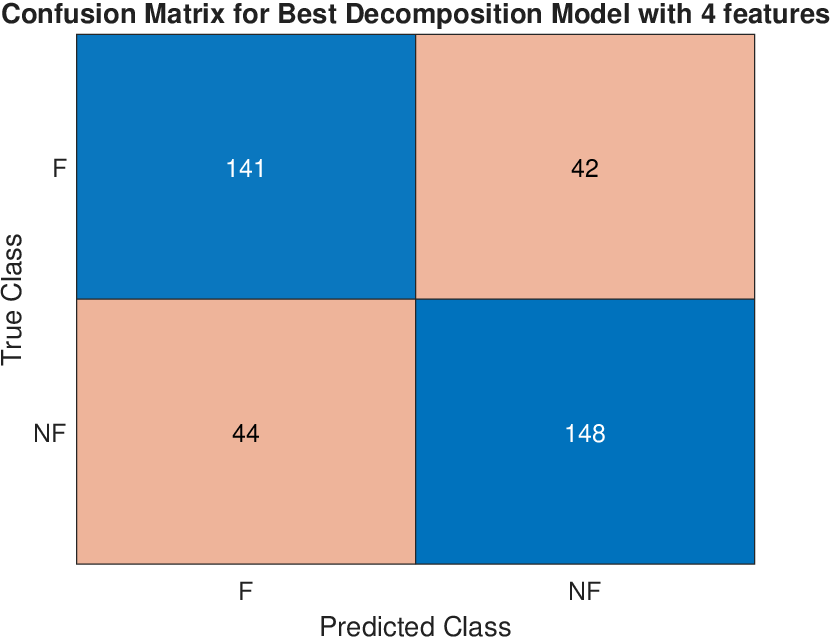}
    \caption{Composite, 4 features: $\bar{\Delta}t^{(r)}_{1}$, $\bar{\delta}t_1$, $\bar{\delta}t_3$, $\bar{\Delta} A^{(c)}_2$}
\end{subfigure}

\caption{Confusion matrices for the top-performing classification models using (a–c) classical features and (d–f) decomposition-based composite features. Columns correspond to models trained with 2, 3, and 4 features, respectively. Values represent the number of breaths classified as fatigued (F) or non-fatigued (NF) across all test folds.}
\label{fig:RQ3_confmat_all}
\end{figure*}

For the composite feature pool, the best two-feature model ($\bar{\Delta} t^{(r)}_1$, $\bar{\delta}t_2$) achieved MCC = 0.482 and $F_1$ = 0.733, representing a 15.0\% increase in MCC relative to the classical model. The best three-feature model ($\bar{\Delta} t^{(c)}_1$, $\bar{\Delta}t^{(r)}_{1}$, $\bar{\Delta}A^{(r)}_4$) achieved MCC = 0.536 and $F_1$ = 0.759, a 28.0\% increase. The best four-feature model ($\bar{\Delta} t^{(r)}_1$, $\bar{\delta}t_1$, $\bar{\delta}t_3$, $\bar{\Delta} A^{(c)}_2$) achieved MCC = 0.541 and $F_1$ = 0.766, a 30.7\% increase compared to the classical model. Confusion matrices for the highest-performing models across both feature pools are shown in Figure~\ref{fig:RQ3_confmat_all}.

\section{Discussion}
\subsection{RQ1 - Can inspiratory airflow be reconstructed at the single-breath level using a small number of time-localized, interpretable components?}
For this paper, we limited the analysis to a maximum of six components. This choice was intentional because reconstruction error was already very small by that point, and the incremental reduction in MSE from adding further components would be negligible. Extending the decomposition beyond four components would increase complexity and reduce interpretability without providing meaningful improvements in reconstruction accuracy. Thus, four components were considered an appropriate balance between accuracy and interpretability

The results demonstrate that individual breaths can be reliably decomposed into a sum of time-localized basis functions, with reconstruction accuracy improving systematically as the number of components increases. Across all conditions, MSE decreased by more than an order of magnitude from 1 to 4 components, confirming that a small number of elements can capture most of the morphological variance in inspiratory airflow. This supports the central assumption that breathing waveforms, though complex, can be represented as the linear superposition of a limited number of physiologically interpretable and time-constrained subcomponents.

Although decompositions were computed up to six components, the improvement in reconstruction accuracy beyond four was minimal. As shown in Tables \ref{tab:mse_noload} and \ref{tab:mse_load}, increasing from four to six components reduced mean squared error by only 0.0007–0.001 under load and by roughly 0.0006 under no-load conditions—less than 0.05\% of total signal variance. In contrast, the first four components accounted for more than 95\% of the reduction in error. This indicates that four components capture nearly all meaningful waveform structure, while additional components contribute only marginal refinements.

Clear differences emerged between basis function families. The Beta model achieved the lowest MSE across both load and no-load conditions, followed by the Half-Sine and Gaussian models. The Beta basis’s high flexibility, governed by its $\alpha$ and $\beta$ shape parameters, allows it to accommodate a wide range of waveform asymmetries. However, that same flexibility complicates interpretation: while amplitude and duration map cleanly onto recognizable physical quantities (flow intensity and temporal extent), the abstract shape parameters lack intuitive physiological meaning. In contrast, the Half-Sine basis achieves a good compromise, sufficiently flexible to approximate non-sinusoidal flow shapes, yet constrained to parameters (amplitude, duration, onset) that correspond directly to observable mechanical or neural modulations in breathing. The Gaussian basis, though mathematically stable, performed worst due to its exponential decay, which limits its ability to reproduce the sharp rises and abrupt flow reversals that characterize inspiratory waveforms under load.

Across all models, MSE values were consistently higher for loaded than no load breaths. This pattern indicates that breathing under inspiratory load becomes more variable and less stereotyped in waveform shape. Physiologically, this increased reconstruction error likely reflects adaptive changes in neuromuscular coordination, as additional muscles are recruited and their timing becomes less regular to overcome the imposed load. In other words, the signal itself becomes less easily described by a small number of simple components. The persistence of low MSE values for both Half-Sine and Beta bases, even under load, suggests that the decomposition remains robust to this added complexity and captures meaningful physiological structure rather than overfitting noise.

Physiologically, limiting the model to three to four components provides a plausible correspondence to the principal muscle groups engaged during inspiration. These include the diaphragm as the primary driver of inspiratory flow, the external intercostals supporting ribcage expansion, and upper-thoracic and accessory muscles such as the parasternal intercostals, scalenes, and sternocleidomastoids, which are recruited under increased load or fatigue. Each fitted component can therefore be interpreted as a temporally distinct contribution from a dominant muscle group or from coordinated activations among groups. Adding additional components primarily subdivides these phases rather than introducing new physiologically meaningful events.

While information criteria such as the Bayesian Information Criterion (BIC) or the Akaike Information Criterion (AIC) could be used to adaptively select the number of components for each breath, this would result in variable component counts across samples, complicating the direct comparison of parameters and derived features between breaths or subjects. For this reason, a fixed model order of four components was adopted to maintain interpretability and cross-subject consistency without sacrificing reconstruction fidelity.

\subsection{RQ2 - Does the proposed parametric decomposition yield precise and repeatable component parameters under moderate noise?} 

The observed differences in decomposition stability across basis function families reveal key trade-offs in how structural constraints, flexibility, and physiological interpretability shape the robustness of model parameters under noisy conditions. The Half-Sine model emerges as the most balanced choice, achieving high reconstruction fidelity, strong resistance to noise-induced drift, and intuitive mapping to physiological events.

The Gaussian model showed the highest parameter stability, with minimal variance in amplitude, width, and onset across noisy decompositions. This is primarily due to its rigid structure: the exponential decay of its tails effectively localizes each component and limits overlap, reducing the solution space available to the optimizer. However, this same rigidity constrains its ability to adapt to the asymmetries commonly found in real inspiratory waveforms. As confirmed in RQ1, the Gaussian yielded the worst reconstruction accuracy (highest MSE), underscoring that stability alone is insufficient if the model lacks the expressive power to capture the underlying physiology.

At the other end, the Beta model delivered excellent reconstruction performance, enabled by its tunable shape parameters \(\alpha\) and \(\beta\), which modulate skewness, peak sharpness, and the locations of the rising and falling phases. This flexibility allows the Beta model to accommodate a wide range of waveform morphologies, including those with delayed or abrupt transitions. However, this comes at a cost: under noise, the \(\alpha\) and \(\beta\) parameters become substantially more variable, especially in higher-order components. This reflects a common pitfall in overparameterized models: excess degrees of freedom introduce interdependencies between parameters, degrading robustness and reproducibility. For physiological modeling applications, such instability undermines interpretability and complicates temporal comparisons.

The Half-Sine model, by contrast, provides an optimal trade-off. It demonstrated low reconstruction error, nearly matching the Beta model, while maintaining moderate and well-bounded parameter variability under noise. Its parameter space (amplitude, duration, and onset time) is compact and directly interpretable: each atom corresponds to a physiologically plausible flow segment with smooth onset and offset. Unlike the Beta model, there are no shape parameters that risk collapsing or overfitting to local fluctuations, and, unlike Gaussians, Half-Sines have bounded support and sharper transitions, allowing them to better capture abrupt changes in the signal without interference from long tails. Although slight increases in variability were observed with higher component counts, this did not substantially degrade performance or interpretability. The decomposition remained stable and understandable, even as complexity increased.

To further isolate and quantify parameter precision independent of physiological variability, a complementary simulation study was conducted using synthetically generated inspiratory waveforms. Each waveform was constructed as the sum of three Half-Sine components with randomized but physiologically plausible amplitudes, durations, and onset times, and subsequently perturbed with noise (SNR = 30 dB). The resulting signals were decomposed using the same Half-Sine optimization framework applied to the experimental data. Across 30 unique simulated breaths and 10 noise realizations per breath, reconstruction error remained low (median MSE = $9.10\times10^{-4}$). More importantly, parameter estimates exhibited tight repeatability under noise, with median standard deviations of onset time of 2.5 ms, 9.4 ms, and 20.6 ms across the three components, and duration standard deviations of approximately 35–39 ms. In contrast, absolute recovery accuracy relative to the known ground truth was more variable, particularly for duration (median MAE = 0.132 s) and amplitude (median MAE = 0.088), reflecting the inherent non-uniqueness of decomposing overlapping components. These results reinforce the interpretation of the Half-Sine model as a precise and consistent parameterization of intrabreath airflow morphology rather than a unique physiological source separation. When viewed alongside the experimental results, the simulation confirms that the stability observed in real data arises from the structure of the model itself and not from coincidental properties of a particular dataset.

From a physiological signal processing standpoint, these findings position the Half-Sine basis as the most practical and robust option. It avoids the over-constrained limitations of the Gaussian and the over-flexible pitfalls of the Beta, offering stable, high-fidelity decompositions with clear physiological meaning. This is especially important for downstream tasks such as feature extraction, condition classification, or temporal tracking of respiratory behavior, where both signal reconstruction and parameter interpretability affect the reliability and generalizability of findings.

It is important to emphasize that the goal of this work is not to outperform existing spectral or adaptive methods on generic reconstruction or classification benchmarks, but to address a fundamentally different modeling objective. Methods such as EMD, wavelet transforms, or dictionary learning are designed to distribute signal energy across basis elements, whereas the present framework is designed to recover explicit, time-localized sub-breath events with well-defined onset, duration, and amplitude. Because these approaches optimize different objective functions and yield qualitatively different outputs, direct numerical comparisons would not meaningfully evaluate the parameter precision and physiological interpretability that motivate the proposed framework.

\subsection{RQ3 - Do sub-breath features derived from the proposed representation capture meaningful adaptations in breathing dynamics that are not accessible through conventional breath-level measures?}

The classification results showed that features derived from the decomposition framework improved discrimination between fatigued and non-fatigued breaths compared with classical respiratory metrics. Across all feature set sizes, models incorporating sub-breath descriptors achieved up to 30\% higher MCC scores, indicating that the temporal and amplitude patterns captured by the decomposition framework encode physiologically meaningful variation rather than noise. In particular, timing-related features such as $\bar{\Delta} t^{(r)}_{j}$ and $\Delta t_{k,m}$ consistently ranked among the most discriminative, whereas purely amplitude-based component features such as the average component–component amplitude difference $\bar{\delta}A_k$ did not appear in the best-performing models. This pattern suggests that cognitive fatigue arising from cognitive-respiratory competition manifests as altered temporal coordination within the inspiratory cycle rather than changes in flow magnitude alone.

These parameters reflect changes in how individual components activate and align with the overall airflow pattern, revealing shifts in the internal timing structure of the breath that are not captured by global measures such as total duration, peak flow, or symmetry index. Such shifts may arise when breathing control transitions from predominantly automatic to requiring increased voluntary modulation and attentional resources, potentially altering the recruitment timing and coordination of inspiratory muscle groups as cognitive fatigue develops.

From a physiological perspective, the decomposition-derived features appear to capture aspects of inspiratory motor control adaptation under cognitive-respiratory competition. When mechanical respiratory load exceeds the capacity for automatic compensation, maintaining adequate ventilation may require increased cortical involvement and volitional control, diverting cognitive resources from concurrent tasks and contributing to cognitive fatigue. This shift from automatic to cognitively-mediated breathing control may manifest as compensatory strategies such as modified phase timing, reduced synchrony between subcomponents, and reallocation of effort between early and late inspiratory phases~\cite{bendittRespiratory2019,patelMethods2022}. The framework's time-localized representation enables these dynamics to be quantified directly, explaining the stronger predictive performance of the composite feature models. These features describe how well the reconstructed components align in time and magnitude with the observed airflow signal, thus reflecting changes in motor coordination or synchronization among inspiratory muscle groups that may accompany increased cognitive control of breathing under dual-task conditions.

This interpretation is consistent with established physiological evidence that respiratory loading induces changes in the synchronization and distribution of inspiratory muscle activation. Evidence demonstrates that during progressive inspiratory loading, the onset of activity in accessory muscles, including the scalene, parasternal intercostal, and sternocleidomastoid, shifts earlier relative to airflow onset, indicating a compensatory reorganization of drive timing across muscle groups~\cite{matsumuraTiming2025}. Similarly, literature reports that under sustained load in COPD, neural respiratory drive (diaphragmatic EMG) increases disproportionately to mechanical output, reflecting a loss of neuroventilatory coupling~\cite{jolleyNeural2015}. These findings show that inspiratory control under load becomes less synchronized and more distributed across muscles. The present findings extend this understanding by demonstrating that when loaded breathing occurs concurrently with a demanding cognitive task, the temporal coordination of breathing components—as captured by the decomposition framework—differs systematically between early (non-fatigued) and late (fatigued) phases of the trial. This suggests that decomposition-based features may be sensitive to the degree of cortical involvement or attentional demand associated with loaded breathing, particularly as cognitive fatigue develops from sustained cognitive-respiratory competition.

Traditional features such as $T_i$, $V_T$, and $\dot V_{\text{peak}}$ capture only the global magnitude and duration of airflow and are insensitive to within-breath coordination changes. In contrast, the sub-breath features derived from the decomposition framework explicitly parameterize the timing, amplitude, and relative alignment of individual components, enabling the detection of fine-scale motor control adaptations that would otherwise be averaged out. The improved classification performance, therefore, reflects the framework's ability to resolve physiologically relevant micro-dynamics within each breath—such as delayed recruitment, altered phasing, or redistribution of effort among early and late inspiratory segments—that may indicate transitions between automatic and cognitively-mediated breathing control strategies as cognitive fatigue develops.

It is important to note that this classification task was used as a proof-of-concept application to evaluate whether features derived from the proposed decomposition framework provide added value for discriminating states of cognitive fatigue from breathing signals. Detecting cognitive fatigue arising from cognitive-respiratory competition using airflow alone is inherently difficult due to the subtlety of the underlying physiological changes and the indirect nature of the relationship between central attentional state and peripheral respiratory motor patterns. Fatigue labels were derived from changes in PVT performance, specifically, the slope of response time across the high-load trial, under the interpretation that vigilance decline during loaded breathing reflects insufficient automatic compensation, consequent competition for cognitive resources, and resulting cognitive fatigue from sustained dual-task demands. As such, this analysis tested whether respiratory dynamics contained sufficient information to reflect cognitive fatigue induced by cognitive-respiratory competition. The task was not intended as a comprehensive solution for cognitive fatigue detection, but rather as one representative example of how the framework can be used to probe latent physiological adaptations in a controlled context.

The analysis focused on the 10 participants who exhibited measurable cognitive fatigue (positive PVT slope) because the goal was to determine whether respiratory airflow alone could capture morphological signatures associated with cognitive-respiratory competition within the same individuals. Including participants whose vigilance remained stable would have diluted this within-subject contrast, as their respiratory patterns likely reflected successful automatic compensation without significant cognitive resource diversion or resulting cognitive fatigue. Breaths within each trial were therefore labeled as "non-fatigued" (early phase, 10–20\%) and "fatigued" (late phase, >90\%) to maximize the distinction between states of minimal and substantial cognitive-respiratory competition. Future work should adopt a continuous modeling approach, such as regression or mixed-effects analysis linking decomposition-derived features directly to PVT slope or reaction-time trajectories, and should also investigate whether similar breathing pattern changes occur in respiratory muscle fatigue to distinguish cognitive fatigue from peripheral fatigue mechanisms.

Overall, these findings demonstrate that the proposed decomposition framework, together with its derived sub-breath feature set, provides a representation of inspiratory intrabreath events that exposes latent temporal coordination and effort redistribution within each breath. These features capture adaptations in respiratory motor control associated with cognitive fatigue arising from cognitive-respiratory competition that are fundamentally inaccessible to conventional breath-level metrics. While classification accuracy was not the primary focus of this work, the results validate the framework's potential for downstream applications in cognitive workload assessment, dual-task performance monitoring, and adaptive control of respiratory interfaces.

\section{Conclusion}
This study introduced a parametric framework for decomposing inspiratory airflow into a small number of time-localized components with explicit timing and amplitude parameters. By representing each breath as a structured sum of constrained waveform primitives, the framework produces a compact, interpretable description of intrabreath airflow that differs fundamentally from spectral, time–frequency, and purely breath-level analyses. Instead of relying on data-adaptive or oscillatory methods, the approach encodes temporal structure directly in the model, which supports consistent reconstruction, parameter estimation, and feature extraction at the level of single breaths.

Three major contributions were introduced in this study. First, we showed that inspiratory airflow can be reconstructed with high accuracy, indicating that complex waveform shapes can be captured using only a few physiologically plausible components. Second, we demonstrate that the proposed model produces stable and repeatable component parameters in the presence of moderate noise, thereby providing the reliability necessary for comparisons across breaths and experimental conditions. Third, the features derived from this decomposition, especially those that describe sub-breath timing, coordination, and alignment, were substantially more sensitive to fatigue-related changes than conventional respiratory measures. This emphasizes the value of resolving internal breath organization rather than relying only on global summary metrics.

The findings support the practical utility of the framework. The consistent advantage of timing- and coordination-based features over amplitude-based descriptors suggests that respiratory adaptation under cognitive and physiological stress is expressed through changes in the organization of inspiratory motor control, rather than through uniform scaling of airflow. By providing a stable coordinate system for these temporal relationships, the framework enables analysis of latent respiratory dynamics that traditional metrics cannot access. This is directly relevant for studies of respiratory control, compensatory breathing strategies, and the coupling between neural drive and mechanical output.

Future work should extend this framework in several directions. One promising application is the early detection and characterization of impending respiratory failure in extreme environments, including hypoxia, hypercapnia, and underwater diving. In these settings, compensatory strategies often appear first as subtle changes in inspiratory timing and coordination, well before large changes in ventilation or airflow amplitude are evident. The proposed decomposition and its sub-breath features, therefore, may provide sensitive markers of inability to compensate prior to ventilatory collapse.

The same logic applies to clinical populations, such as individuals with chronic respiratory disease or neuromuscular disorders, where delayed recruitment, altered activation patterns, and loss of synchrony among inspiratory muscles are key features of disease progression. By quantifying intrabreath timing and the redistribution of effort using airflow alone, the framework offers a noninvasive way to characterize these changes. Integrating this approach with electromyography or measures of neural respiratory drive, including respiratory-related evoked potentials, would further strengthen physiological interpretation by linking component timing directly to underlying motor activation. Such multimodal studies could clarify how decomposition-derived components relate to neural drive, muscle recruitment, and mechanical output, and help establish the framework as a tool for early detection, monitoring, and mechanistic investigation of respiratory dysfunction.

\section*{Acknowledgment}
This work was supported by the Office of Naval Research (grant number N00014-22-1-2653).

\ifCLASSOPTIONcaptionsoff
  \newpage
\fi
\bibliographystyle{ieeetr}
\bibliography{bibtex/bib/references}

@article{11283643,
	title = {Attractor reconstruction of breathing dynamics: {Characterising} respiratory dysfunction in {COPD}},
	volume = {29},
	doi = {10.1109/JBHI.2025.3605799},
	number = {12},
	journal = {IEEE Journal of Biomedical and Health Informatics},
	author = {Chanchotisatien, Passara and Arvind, DK},
	year = {2025},
	pages = {8687--8694},
}

@article{9372748,
	title = {{CNN}-{MoE} based framework for classification of respiratory anomalies and lung disease detection},
	volume = {25},
	doi = {10.1109/JBHI.2021.3064237},
	number = {8},
	journal = {IEEE Journal of Biomedical and Health Informatics},
	author = {Pham, Lam and Phan, Huy and Palaniappan, Ramaswamy and Mertins, Alfred and McLoughlin, Ian},
	year = {2021},
	pages = {2938--2947},
}

@article{7101812,
	title = {Estimation of respiratory rate from photoplethysmographic imaging videos compared to pulse oximetry},
	volume = {19},
	doi = {10.1109/JBHI.2015.2429746},
	number = {4},
	journal = {IEEE Journal of Biomedical and Health Informatics},
	author = {Karlen, Walter and Garde, Ainara and Myers, Dorothy and Scheffer, Cornie and Ansermino, J Mark and Dumont, Guy A},
	year = {2015},
	pages = {1331--1338},
}

@article{10904160,
	title = {Respiratory anomaly and disease detection using multi-level temporal convolutional networks},
	volume = {29},
	doi = {10.1109/JBHI.2025.3545156},
	number = {7},
	journal = {IEEE Journal of Biomedical and Health Informatics},
	author = {Le, Kim-Ngoc T. and Byun, Gyurin and Raza, Syed M. and Le, Duc-Tai and Choo, Hyunseung},
	year = {2025},
	pages = {4834--4846},
}

@article{7452349,
	title = {Inclusion of respiratory frequency information in heart rate variability analysis for stress assessment},
	volume = {20},
	doi = {10.1109/JBHI.2016.2553578},
	number = {4},
	journal = {IEEE Journal of Biomedical and Health Informatics},
	author = {Hernando, Alberto and Lázaro, Jesús and Gil, Eduardo and Arza, Adriana and Garzón, Jorge Mario and López-Antón, Raúl and de la Cámara, Concepción and Laguna, Pablo and Aguiló, Jordi and Bailón, Raquel},
	year = {2016},
	pages = {1016--1025},
}

@article{7873222,
	title = {Ensemble empirical mode decomposition with principal component analysis: a novel approach for extracting respiratory rate and heart rate from photoplethysmographic signal},
	volume = {22},
	doi = {10.1109/JBHI.2017.2679108},
	number = {3},
	journal = {IEEE Journal of Biomedical and Health Informatics},
	author = {Motin, Mohammod Abdul and Karmakar, Chandan Kumar and Palaniswami, Marimuthu},
	year = {2018},
	pages = {766--774},
}

@article{9591404,
	title = {Comprehensive analysis system for automated respiratory cycle segmentation and crackle peak detection},
	volume = {26},
	doi = {10.1109/JBHI.2021.3123353},
	number = {4},
	journal = {IEEE Journal of Biomedical and Health Informatics},
	author = {McLane, Ian and Lauwers, Eline and Stas, Toon and Busch-Vishniac, Ilene and Ides, Kris and Verhulst, Stijn and Steckel, Jan},
	year = {2022},
	pages = {1847--1860},
}

@article{7021891,
	title = {Automatic differentiation of normal and continuous adventitious respiratory sounds using ensemble empirical mode decomposition and instantaneous frequency},
	volume = {20},
	doi = {10.1109/JBHI.2015.2396636},
	number = {2},
	journal = {IEEE Journal of Biomedical and Health Informatics},
	author = {Lozano, Manuel and Fiz, José Antonio and Jané, Raimon},
	year = {2016},
	pages = {486--497},
}

@article{9373894,
	title = {Automatic respiratory event scoring in obstructive sleep apnea using a long short-term memory neural network},
	volume = {25},
	doi = {10.1109/JBHI.2021.3064694},
	number = {8},
	journal = {IEEE Journal of Biomedical and Health Informatics},
	author = {Nikkonen, Sami and Korkalainen, Henri and Leino, Akseli and Myllymaa, Sami and Duce, Brett and Leppänen, Timo and Töyräs, Juha},
	year = {2021},
	pages = {2917--2927},
}

@article{9310306,
	title = {A lightweight {CNN} model for detecting respiratory diseases from lung auscultation sounds using {EMD}-{CWT}-based hybrid scalogram},
	volume = {25},
	doi = {10.1109/JBHI.2020.3048006},
	number = {7},
	journal = {IEEE Journal of Biomedical and Health Informatics},
	author = {Shuvo, Samiul Based and Ali, Shams Nafisa and Swapnil, Soham Irtiza and Hasan, Taufiq and Bhuiyan, Mohammed Imamul Hassan},
	year = {2021},
	pages = {2595--2603},
}

@article{8688505,
	title = {Non-contact sleep stage detection using canonical correlation analysis of respiratory sound},
	volume = {24},
	doi = {10.1109/JBHI.2019.2910566},
	number = {2},
	journal = {IEEE Journal of Biomedical and Health Informatics},
	author = {Xue, Biao and Deng, Boya and Hong, Hong and Wang, Zhiyong and Zhu, Xiaohua and Feng, David Dagan},
	year = {2020},
	pages = {614--625},
}

@article{10521732,
	title = {Quantifying posttraumatic stress disorder symptoms during traumatic memories using interpretable markers of respiratory variability},
	volume = {28},
	doi = {10.1109/JBHI.2024.3397589},
	number = {8},
	journal = {IEEE Journal of Biomedical and Health Informatics},
	author = {Gazi, Asim H. and Sanchez-Perez, Jesus Antonio and Saks, Georgia L. and Alday, Erick A. Perez and Haffar, Ammer and Ahmed, Hashir and Herraka, Duaa and Tarlapally, Nitya and Smith, Nicholas L. and Bremner, J. Douglas and Shah, Amit J. and Inan, Omer T. and Vaccarino, Viola},
	year = {2024},
	pages = {4912--4924},
}

@article{bendittRespiratory2019,
	title = {Respiratory {Care} of {Patients} {With} {Neuromuscular} {Disease}},
	volume = {64},
	copyright = {https://www.liebertpub.com/nv/resources-tools/text-and-data-mining-policy/121/},
	issn = {0020-1324},
	url = {https://www.liebertpub.com/doi/10.4187/respcare.06827},
	doi = {10.4187/respcare.06827},
	language = {en},
	number = {6},
	urldate = {2025-12-29},
	journal = {Respiratory Care},
	author = {Benditt, Joshua O},
	month = jun,
	year = {2019},
	pages = {679--688},
}

@article{patelMethods2022,
	title = {Methods and {Applications} in {Respiratory} {Physiology}: {Respiratory} {Mechanics}, {Drive} and {Muscle} {Function} in {Neuromuscular} and {Chest} {Wall} {Disorders}},
	volume = {13},
	issn = {1664-042X},
	shorttitle = {Methods and {Applications} in {Respiratory} {Physiology}},
	url = {https://pmc.ncbi.nlm.nih.gov/articles/PMC9237333/},
	doi = {10.3389/fphys.2022.838414},
	urldate = {2025-12-29},
	journal = {Frontiers in Physiology},
	author = {Patel, Nina and Chong, Kelvin and Baydur, Ahmet},
	month = jun,
	year = {2022},
	pages = {838414},
}

@article{huangempirical1998,
	title = {The empirical mode decomposition and the {Hilbert} spectrum for nonlinear and non-stationary time series analysis},
	volume = {454},
	issn = {1364-5021},
	url = {https://doi.org/10.1098/rspa.1998.0193},
	doi = {10.1098/rspa.1998.0193},
	number = {1971},
	urldate = {2025-12-28},
	journal = {Proceedings of the Royal Society A: Mathematical, Physical and Engineering Sciences},
	author = {Huang, Norden E. and Shen, Zheng and Long, Steven R. and Wu, Manli C. and Shih, Hsing H. and Zheng, Quanan and Yen, Nai-Chyuan and Tung, Chi Chao and Liu, Henry H.},
	month = mar,
	year = {1998},
	pages = {903--995},
}

@article{milic-emiliDrive1976,
	title = {Drive and timing components of ventilation},
	volume = {70},
	issn = {0012-3692},
	doi = {10.1378/chest.70.1_supplement.131},
	language = {eng},
	number = {1 Suppl},
	journal = {Chest},
	author = {Milic-Emili, J. and Grunstein, M. M.},
	month = jul,
	year = {1976},
	pages = {131--133},
}

@article{benchetritBreathing2000,
	title = {Breathing pattern in humans: diversity and individuality},
	volume = {122},
	issn = {0034-5687},
	shorttitle = {Breathing pattern in humans},
	doi = {10.1016/s0034-5687(00)00154-7},
	language = {eng},
	number = {2-3},
	journal = {Respiration Physiology},
	author = {Benchetrit, G.},
	month = sep,
	year = {2000},
	pages = {123--129},
}

@inproceedings{yadavMachine2020,
	title = {Machine {Learning} {Based} {Automatic} {Classification} of {Respiratory} {Signals} using {Wavelet} {Transform}},
	url = {https://ieeexplore.ieee.org/abstract/document/9163565},
	doi = {10.1109/TSP49548.2020.9163565},
	urldate = {2025-12-18},
	booktitle = {2020 43rd {International} {Conference} on {Telecommunications} and {Signal} {Processing} ({TSP})},
	author = {Yadav, Anjali and Dutta, Malay Kishore and Prinosil, Jiri},
	month = jul,
	year = {2020},
	pages = {545--549},
}

@inproceedings{kimSignal2001,
	title = {Signal processing using {Fourier} \& wavelet transform for pulse oximetry},
	volume = {2},
	url = {https://ieeexplore.ieee.org/abstract/document/970957},
	doi = {10.1109/CLEOPR.2001.970957},
	urldate = {2025-12-18},
	booktitle = {Technical {Digest}. {CLEO}/{Pacific} {Rim} 2001. 4th {Pacific} {Rim} {Conference} on {Lasers} and {Electro}-{Optics} ({Cat}. {No}.{01TH8557})},
	author = {Kim, J.M. and Kim, S.H. and Lee, D.J. and Lim, H.S.},
	month = jul,
	year = {2001},
	pages = {II--II},
}

@article{mallattheory1989,
	title = {A theory for multiresolution signal decomposition: {The} wavelet representation},
	volume = {11},
	number = {7},
	journal = {IEEE Transactions on Pattern Analysis and Machine Intelligence},
	publisher = {IEEE},
	author = {Mallat, Stéphane},
	year = {1989},
	pages = {674--693},
}

@article{zechmanEffect1977,
	title = {Effect of chest cage restriction on perception of added airflow resistance},
	volume = {31},
	url = {https://doi.org/10.1016/0034-5687(77)90066-4},
	doi = {10.1016/0034-5687(77)90066-4},
	number = {1},
	journal = {Respiration Physiology},
	publisher = {Elsevier BV},
	author = {Zechman, F.W. and Wiley, R.L.},
	month = sep,
	year = {1977},
	pages = {71--79},
}

@article{gaborTheory1946,
	title = {Theory of communication},
	volume = {93},
	number = {26},
	journal = {Journal of the Institution of Electrical Engineers-Part III: Radio and Communication Engineering},
	publisher = {IET},
	author = {Gabor, Dennis},
	year = {1946},
	pages = {429--457},
}

@incollection{milic-emiliRelationship2011,
	title = {Relationship between neuromuscular respiratory drive and ventilatory output},
	isbn = {978-0-470-65071-4},
	url = {https://onlinelibrary.wiley.com/doi/abs/10.1002/cphy.cp030335},
	booktitle = {Comprehensive physiology},
	publisher = {John Wiley \& Sons, Ltd},
	author = {Milic-Emili, J. and Zin, W. A.},
	year = {2011},
	doi = {https://doi.org/10.1002/cphy.cp030335},
	note = {tex.eprint: https://onlinelibrary.wiley.com/doi/pdf/10.1002/cphy.cp030335},
	pages = {631--646},
}

@article{hessNeural2013,
	title = {Neural {Mechanisms} {Underlying} {Breathing} {Complexity}},
	volume = {8},
	issn = {1932-6203},
	url = {https://journals.plos.org/plosone/article?id=10.1371/journal.pone.0075740},
	doi = {10.1371/journal.pone.0075740},
	language = {en},
	number = {10},
	urldate = {2022-10-20},
	journal = {PLOS ONE},
	publisher = {Public Library of Science},
	author = {Hess, Agathe and Yu, Lianchun and Klein, Isabelle and Mazancourt, Marine De and Jebrak, Gilles and Mal, Hervé and Brugière, Olivier and Fournier, Michel and Courbage, Maurice and Dauriat, Gaelle and Schouman-Clayes, Elisabeth and Clerici, Christine and Mangin, Laurence},
	month = oct,
	year = {2013},
	pages = {e75740},
}

@article{ribeirorodriguesChest2025,
	title = {Chest wall restriction device for modeling respiratory challenges and dysfunction},
	volume = {3},
	issn = {2813-687X},
	url = {https://www.frontiersin.org/journals/medical-engineering/articles/10.3389/fmede.2025.1560136/full},
	doi = {10.3389/fmede.2025.1560136},
	language = {English},
	urldate = {2025-09-30},
	journal = {Frontiers in Medical Engineering},
	publisher = {Frontiers},
	author = {Ribeiro Rodrigues, Victoria and Mejia, Lizuannette and Zucchi, Rafael G. and Davenport, Paul W. and Napoli, Nicholas J.},
	month = may,
	year = {2025},
}

@article{vicarioNoninvasive2016,
	title = {Noninvasive {Estimation} of {Respiratory} {Mechanics} in {Spontaneously} {Breathing} {Ventilated} {Patients}: {A} {Constrained} {Optimization} {Approach}},
	volume = {63},
	issn = {1558-2531},
	shorttitle = {Noninvasive {Estimation} of {Respiratory} {Mechanics} in {Spontaneously} {Breathing} {Ventilated} {Patients}},
	url = {https://ieeexplore.ieee.org/abstract/document/7214248},
	doi = {10.1109/TBME.2015.2470641},
	number = {4},
	urldate = {2025-10-20},
	journal = {IEEE Transactions on Biomedical Engineering},
	author = {Vicario, Francesco and Albanese, Antonio and Karamolegkos, Nikolaos and Wang, Dong and Seiver, Adam and Chbat, Nicolas W.},
	month = apr,
	year = {2016},
	pages = {775--787},
}

@article{onimaruNovel2003,
	chapter = {ARTICLE},
	title = {A {Novel} {Functional} {Neuron} {Group} for {Respiratory} {Rhythm} {Generation} in the {Ventral} {Medulla}},
	volume = {23},
	copyright = {Copyright © 2003 Society for Neuroscience},
	issn = {0270-6474, 1529-2401},
	url = {https://www.jneurosci.org/content/23/4/1478},
	doi = {10.1523/JNEUROSCI.23-04-01478.2003},
	language = {en},
	number = {4},
	urldate = {2025-10-20},
	journal = {Journal of Neuroscience},
	publisher = {Society for Neuroscience},
	author = {Onimaru, Hiroshi and Homma, Ikuo},
	month = feb,
	year = {2003},
	pages = {1478--1486},
}

@article{fiammaEffects2007,
	title = {Effects of hypercapnia and hypocapnia on ventilatory variability and the chaotic dynamics of ventilatory flow in humans},
	volume = {292},
	issn = {0363-6119},
	url = {https://journals.physiology.org/doi/full/10.1152/ajpregu.00792.2006},
	doi = {10.1152/ajpregu.00792.2006},
	number = {5},
	urldate = {2025-10-20},
	journal = {American Journal of Physiology-Regulatory, Integrative and Comparative Physiology},
	publisher = {American Physiological Society},
	author = {Fiamma, Marie-Noëlle and Straus, Christian and Thibault, Sylvain and Wysocki, Marc and Baconnier, Pierre and Similowski, Thomas},
	month = may,
	year = {2007},
	pages = {R1985--R1993},
}

@article{ciolekAutomated2015,
	title = {Automated {Detection} of {Sleep} {Apnea} and {Hypopnea} {Events} {Based} on {Robust} {Airflow} {Envelope} {Tracking} in the {Presence} of {Breathing} {Artifacts}},
	volume = {19},
	issn = {2168-2208},
	url = {https://ieeexplore.ieee.org/abstract/document/6820757},
	doi = {10.1109/JBHI.2014.2325997},
	number = {2},
	urldate = {2025-10-20},
	journal = {IEEE Journal of Biomedical and Health Informatics},
	author = {Ciołek, Marcin and Niedźwiecki, Maciej and Sieklicki, Stefan and Drozdowski, Jacek and Siebert, Janusz},
	month = mar,
	year = {2015},
	pages = {418--429},
}

@article{harberPhysiologic1982,
	title = {Physiologic {Effects} of {Respirator} {Dead} {Space} and {Resistance} {Loading}:},
	volume = {24},
	issn = {1076-2752},
	shorttitle = {Physiologic {Effects} of {Respirator} {Dead} {Space} and {Resistance} {Loading}},
	url = {http://journals.lww.com/00043764-198209000-00015},
	doi = {10.1097/00043764-198209000-00015},
	language = {en},
	number = {9},
	urldate = {2025-10-07},
	journal = {Journal of Occupational and Environmental Medicine},
	author = {Harber, Philip and Tamimie, R Joseph and Bhattacharya, Amit and Barber, Melvin},
	month = sep,
	year = {1982},
	pages = {681--684},
}

@article{grassmannRespiratory2016a,
	title = {Respiratory {Changes} in {Response} to {Cognitive} {Load}: {A} {Systematic} {Review}},
	volume = {2016},
	copyright = {http://creativecommons.org/licenses/by/4.0/},
	issn = {2090-5904, 1687-5443},
	shorttitle = {Respiratory {Changes} in {Response} to {Cognitive} {Load}},
	url = {http://www.hindawi.com/journals/np/2016/8146809/},
	doi = {10.1155/2016/8146809},
	language = {en},
	urldate = {2025-10-07},
	journal = {Neural Plasticity},
	author = {Grassmann, Mariel and Vlemincx, Elke and Von Leupoldt, Andreas and Mittelstädt, Justin M. and Van Den Bergh, Omer},
	year = {2016},
	pages = {1--16},
}

@article{romerExerciseinduced2008,
	title = {Exercise-induced respiratory muscle fatigue: implications for performance},
	volume = {104},
	issn = {8750-7587, 1522-1601},
	shorttitle = {Exercise-induced respiratory muscle fatigue},
	url = {https://www.physiology.org/doi/10.1152/japplphysiol.01157.2007},
	doi = {10.1152/japplphysiol.01157.2007},
	language = {en},
	number = {3},
	urldate = {2025-10-07},
	journal = {Journal of Applied Physiology},
	author = {Romer, Lee M. and Polkey, Michael I.},
	month = mar,
	year = {2008},
	pages = {879--888},
}

@article{mckenzieRespiratory2012,
	title = {Respiratory physiology: adaptations to high-level exercise},
	volume = {46},
	issn = {0306-3674, 1473-0480},
	shorttitle = {Respiratory physiology},
	url = {https://bjsm.bmj.com/lookup/doi/10.1136/bjsports-2011-090824},
	doi = {10.1136/bjsports-2011-090824},
	language = {en},
	number = {6},
	urldate = {2025-10-07},
	journal = {British Journal of Sports Medicine},
	author = {McKenzie, Donald C},
	month = may,
	year = {2012},
	pages = {381--384},
}

@article{jolleyNeural2015,
	chapter = {Original Articles},
	title = {Neural respiratory drive and breathlessness in {COPD}},
	volume = {45},
	issn = {0903-1936, 1399-3003},
	url = {https://publications.ersnet.org/content/erj/45/2/355},
	doi = {10.1183/09031936.00063014},
	language = {en},
	number = {2},
	urldate = {2025-10-06},
	journal = {European Respiratory Journal},
	publisher = {European Respiratory Society},
	author = {Jolley, Caroline J. and Luo, Yuanming M. and Steier, Joerg and Rafferty, Gerrard F. and Polkey, Michael I. and Moxham, John},
	month = jan,
	year = {2015},
	pages = {355--364},
}

@article{matsumuraTiming2025,
	title = {Timing of activation of different inspiratory muscles during incremental inspiratory loading in healthy adults: {A} cross-sectional study},
	volume = {61},
	issn = {1205-9838},
	shorttitle = {Timing of activation of different inspiratory muscles during incremental inspiratory loading in healthy adults},
	url = {https://pmc.ncbi.nlm.nih.gov/articles/PMC12357601/},
	doi = {10.29390/001c.143022},
	urldate = {2025-10-06},
	journal = {Canadian Journal of Respiratory Therapy: CJRT = Revue Canadienne de la Thérapie Respiratoire : RCTR},
	author = {Matsumura, Umi and Rodrigues, Antenor and Mori, Tamires and Rassam, Peter and Van Hollebeke, Marine and Rozenberg, Dmitry and Brochard, Laurent and Goligher, Ewan C and Roblyer, Darren and Reid, W Darlene},
	year = {2025},
	pages = {191--205},
}

@article{sommermeyerDetection,
	title = {Detection of {Sleep} {Disordered} {Breathing} and {Its} {Central}/{Obstructive} {Character} {Using} {Nasal} {Cannula} and {Finger} {Pulse} {Oximeter}},
	volume = {08},
	url = {https://jcsm.aasm.org/doi/10.5664/jcsm.2148},
	doi = {10.5664/jcsm.2148},
	number = {05},
	urldate = {2025-10-21},
	journal = {Journal of Clinical Sleep Medicine},
	publisher = {American Academy of Sleep Medicine},
	author = {Sommermeyer, Dirk and Zou, Ding and Grote, Ludger and Hedner, Jan},
	pages = {527--533},
}

@article{salisburyRapid2007,
	title = {Rapid screening test for sleep apnea using a nonlinear and nonstationary signal processing technique},
	volume = {29},
	issn = {1350-4533},
	url = {https://www.sciencedirect.com/science/article/pii/S1350453306001081},
	doi = {10.1016/j.medengphy.2006.05.013},
	number = {3},
	urldate = {2025-10-20},
	journal = {Medical Engineering \& Physics},
	author = {Salisbury, John I. and Sun, Ying},
	month = apr,
	year = {2007},
	pages = {336--343},
}

@inproceedings{robertsonEMD2007,
	title = {{EMD} and {PCA} for the {Prediction} of {Sleep} {Apnoea}: {A} {Comparative} {Study}},
	issn = {2162-7843},
	shorttitle = {{EMD} and {PCA} for the {Prediction} of {Sleep} {Apnoea}},
	url = {https://ieeexplore.ieee.org/document/4458166},
	doi = {10.1109/ISSPIT.2007.4458166},
	urldate = {2025-10-20},
	booktitle = {2007 {IEEE} {International} {Symposium} on {Signal} {Processing} and {Information} {Technology}},
	author = {Robertson, H.J. and Soraghan, J.J. and Idzikowski, C. and Conway, B.A.},
	month = dec,
	year = {2007},
	pages = {419--424},
}

@article{serna-pascualNovel2023,
	title = {Novel breathing pattern analysis: {Symmetric} {Projection} {Attractor} {Reconstruction} improves identification of impending {COPD} re-exacerbations – a retrospective cohort analysis},
	volume = {9},
	issn = {2312-0541},
	shorttitle = {Novel breathing pattern analysis},
	url = {https://publications.ersnet.org/lookup/doi/10.1183/23120541.00164-2023},
	doi = {10.1183/23120541.00164-2023},
	language = {en},
	number = {4},
	urldate = {2025-10-20},
	journal = {ERJ Open Research},
	author = {Serna-Pascual, Miquel and D'Cruz, Rebecca F. and Volovaya, Maria and Jolley, Caroline J. and Hart, Nicholas and Rafferty, Gerrard F. and Steier, Joerg and Aston, Philip J. and Nandi, Manasi},
	month = jul,
	year = {2023},
	pages = {00164--2023},
}

@article{daubechiesOrthonormal1988,
	title = {Orthonormal bases of compactly supported wavelets},
	volume = {41},
	number = {7},
	journal = {Communications on Pure and Applied Mathematics},
	publisher = {Wiley Online Library},
	author = {Daubechies, Ingrid},
	year = {1988},
	pages = {909--996},
}

@article{allenShort1977,
	title = {Short term spectral analysis, synthesis, and modification by discrete {Fourier} transform},
	volume = {25},
	issn = {0096-3518},
	url = {http://dx.doi.org/10.1109/TASSP.1977.1162950},
	doi = {10.1109/tassp.1977.1162950},
	number = {3},
	journal = {IEEE Transactions on Acoustics, Speech, and Signal Processing},
	publisher = {Institute of Electrical and Electronics Engineers (IEEE)},
	author = {Allen, J.},
	month = jun,
	year = {1977},
	pages = {235--238},
}

@book{napoliCharacterizing2018,
	title = {Characterizing uncertainty in sensor fusion to improve predictive models},
	publisher = {Online Archive of University of Virginia},
	author = {Napoli, N.J.},
	year = {2018},
	note = {pages: 1-201},
}

@article{Byrd1999,
	title = {An interior point algorithm for large-scale nonlinear programming},
	volume = {9},
	issn = {1095-7189},
	url = {http://dx.doi.org/10.1137/S1052623497325107},
	doi = {10.1137/s1052623497325107},
	number = {4},
	journal = {SIAM Journal on Optimization},
	publisher = {Society for Industrial \& Applied Mathematics (SIAM)},
	author = {Byrd, Richard H. and Hribar, Mary E. and Nocedal, Jorge},
	month = jan,
	year = {1999},
	pages = {877--900},
}

@article{Byrd2000,
	title = {A trust region method based on interior point techniques for nonlinear programming},
	volume = {89},
	issn = {0025-5610},
	url = {http://dx.doi.org/10.1007/PL00011391},
	doi = {10.1007/pl00011391},
	number = {1},
	journal = {Mathematical Programming},
	publisher = {Springer Science and Business Media LLC},
	author = {Byrd, Richard H. and Gilbert, Jean Charles and Nocedal, Jorge},
	month = nov,
	year = {2000},
	pages = {149--185},
}

@article{Waltz2005,
	title = {An interior algorithm for nonlinear optimization that combines line search and trust region steps},
	volume = {107},
	issn = {1436-4646},
	url = {http://dx.doi.org/10.1007/s10107-004-0560-5},
	doi = {10.1007/s10107-004-0560-5},
	number = {3},
	journal = {Mathematical Programming},
	publisher = {Springer Science and Business Media LLC},
	author = {Waltz, R.A. and Morales, J.L. and Nocedal, J. and Orban, D.},
	month = nov,
	year = {2005},
	pages = {391--408},
}

@article{barroso-garciaWavelet2021,
	title = {Wavelet {Analysis} of {Overnight} {Airflow} to {Detect} {Obstructive} {Sleep} {Apnea} in {Children}},
	volume = {21},
	issn = {1424-8220},
	url = {https://www.mdpi.com/1424-8220/21/4/1491},
	doi = {10.3390/s21041491},
	language = {en},
	number = {4},
	urldate = {2023-07-06},
	journal = {Sensors},
	author = {Barroso-García, Verónica and Gutiérrez-Tobal, Gonzalo C. and Gozal, David and Vaquerizo-Villar, Fernando and Álvarez, Daniel and Del Campo, Félix and Kheirandish-Gozal, Leila and Hornero, Roberto},
	month = feb,
	year = {2021},
	pages = {1491},
}

@book{cohen1995,
	address = {Englewood Cliffs, NJ},
	title = {Time-frequency analysis},
	publisher = {Prentice Hall},
	author = {Cohen, Leon},
	year = {1995},
}

@book{bracewell2000,
	address = {New York},
	edition = {3rd},
	title = {The fourier transform and its applications},
	publisher = {McGraw-Hill},
	author = {Bracewell, Ronald N.},
	year = {2000},
}

@inproceedings{ciolekanalysis2010,
	address = {Rzeszow, Poland},
	title = {The analysis of patients' airflow with respect to early detection of sleep apnea},
	isbn = {978-1-4244-7560-5},
	url = {http://ieeexplore.ieee.org/document/5514560/},
	doi = {10.1109/HSI.2010.5514560},
	urldate = {2023-07-20},
	booktitle = {3rd {International} {Conference} on {Human} {System} {Interaction}},
	publisher = {IEEE},
	author = {Ciolek, Marcin and Sieklicki, Stefan and Drozdowski, Jacek},
	month = may,
	year = {2010},
	pages = {241--245},
}

@article{grayRespiration1951,
	title = {Respiration},
	volume = {13},
	number = {2},
	journal = {Annual Review Physiology},
	author = {Gray, J.S. and Grodin, F.S.},
	year = {1951},
	pages = {217--232},
}

@article{satoMethods2001,
	title = {Methods for averaging irregular respiratory flow profiles in awake humans},
	volume = {90},
	number = {2},
	journal = {Journal of Applied Physiology},
	author = {Sato, Jiro and Robbins, Peter A.},
	year = {2001},
	pages = {705--712},
}

@article{benchetritIndividuality1989,
	title = {Individuality of breathing patterns in adults assessed over time},
	volume = {75},
	number = {2},
	journal = {Respiration Physiology},
	author = {Benchetrit, G. and Shea, S.A. and Dinh, T.Pham and Bodocco, S. and Baconnier, P. and Guz, A.},
	year = {1989},
	pages = {199 -- 209},
}

@article{abboudFrequency1986,
	title = {Frequency and time domain analysis of airflow breath patterns in patients with chronic obstructive airway disease},
	volume = {19},
	number = {3},
	journal = {Computers and Biomedical Research},
	author = {Abboud, Shimon and Bruderman, Israel and Sadeh, Dror},
	year = {1986},
	pages = {266 -- 273},
}

@article{otisEffect1949,
	title = {Effect of {Gas} {Density} on {Resistance} to {Respiratory} {Gas} {Flow} in {Man}},
	volume = {2},
	issn = {8750-7587, 1522-1601},
	url = {http://www.physiology.org/doi/10.1152/jappl.1949.2.6.300},
	doi = {10.1152/jappl.1949.2.6.300},
	language = {en},
	number = {6},
	urldate = {2022-10-25},
	journal = {Journal of Applied Physiology},
	author = {Otis, Arthur B. and Bembower, Weldon C.},
	month = dec,
	year = {1949},
	pages = {300--306},
}

@article{emintaglukClassification2010,
	title = {Classıfıcation of sleep apnea by using wavelet transform and artificial neural networks},
	volume = {37},
	issn = {09574174},
	url = {https://linkinghub.elsevier.com/retrieve/pii/S0957417409006095},
	doi = {10.1016/j.eswa.2009.06.049},
	language = {en},
	number = {2},
	urldate = {2023-07-06},
	journal = {Expert Systems with Applications},
	author = {Emin Tagluk, M. and Akin, Mehmet and Sezgin, Necmettin},
	month = mar,
	year = {2010},
	pages = {1600--1607},
}

@article{napoliCharacterizing2022,
	title = {Characterizing and {Modeling} {Breathing} {Dynamics}: {Flow} {Rate}, {Rhythm}, {Period}, and {Frequency}},
	volume = {12},
	issn = {1664-042X},
	shorttitle = {Characterizing and {Modeling} {Breathing} {Dynamics}},
	url = {https://www.frontiersin.org/articles/10.3389/fphys.2021.772295},
	urldate = {2022-10-25},
	journal = {Frontiers in Physiology},
	author = {Napoli, Nicholas J. and Rodrigues, Victoria R. and Davenport, Paul W.},
	year = {2022},
}

@article{gardeBreathing2010,
	title = {Breathing {Pattern} {Characterization} in {Chronic} {Heart} {Failure} {Patients} {Using} the {Respiratory} {Flow} {Signal}},
	volume = {38},
	issn = {0090-6964, 1573-9686},
	url = {http://link.springer.com/10.1007/s10439-010-0109-0},
	doi = {10.1007/s10439-010-0109-0},
	language = {en},
	number = {12},
	urldate = {2023-07-06},
	journal = {Annals of Biomedical Engineering},
	author = {Garde, A. and Sörnmo, L. and Jané, R. and Giraldo, B. F.},
	month = dec,
	year = {2010},
	pages = {3572--3580},
}

@article{jimenez-garciaAssessment2020,
	title = {Assessment of {Airflow} and {Oximetry} {Signals} to {Detect} {Pediatric} {Sleep} {Apnea}-{Hypopnea} {Syndrome} {Using} {AdaBoost}},
	volume = {22},
	issn = {1099-4300},
	url = {https://www.mdpi.com/1099-4300/22/6/670},
	doi = {10.3390/e22060670},
	language = {en},
	number = {6},
	urldate = {2023-07-06},
	journal = {Entropy},
	author = {Jiménez-García, Jorge and Gutiérrez-Tobal, Gonzalo C. and García, María and Kheirandish-Gozal, Leila and Martín-Montero, Adrián and Álvarez, Daniel and Del Campo, Félix and Gozal, David and Hornero, Roberto},
	month = jun,
	year = {2020},
	pages = {670},
}

@article{bachyprogram1986,
	title = {A program for cycle-by-cycle shape analysis of biological rhythms. {Application} to respiratory rhythm},
	volume = {23},
	issn = {0169-2607},
	number = {3},
	journal = {Computer Methods and Programs in Biomedicine},
	author = {Bachy, Jean-Pierre and Eberhard, Andre and Baconnier, Pierre and Benchetrit, Gila},
	year = {1986},
	pages = {297 -- 307},
}

@article{RODRIGUES2022,
	title = {Exploring inspiratory occlusion metrics to assess respiratory drive in patients under acute intermittent hypoxia},
	volume = {304},
	journal = {Respiratory Physiology \& Neurobiology},
	author = {Rodrigues, Victoria R. and Olsen, Wendy L. and Sajjadi, Elaheh and Smith, Barbara K. and Napoli, Nicholas J.},
	year = {2022},
	pages = {e103922},
}

\end{document}